\documentclass[showpacs,preprintnumbers,amsmath,amssymb]{revtex4}

\usepackage{graphicx}
\usepackage{bm}
\usepackage{color}

\newcommand{\bq}{\begin{equation}}
\newcommand{\eq}{\end{equation}}
\newcommand{\bqa}{\begin{eqnarray}}
\newcommand{\eqa}{\end{eqnarray}}
\newcommand{\nn}{\nonumber \\}

\def\be     {\begin{equation}}
\def\ee     {\end{equation}}
\def\bea        {\begin{eqnarray}}
\def\eea        {\end{eqnarray}}
\def\bnn    {\begin{eqnarray*}}
\def\enn    {\end{eqnarray*}}

\begin{document}

\title{Emergent dual holographic description for interacting Dirac fermions in the large $N$ limit}
\author{Ki-Seok Kim}
\affiliation{Department of Physics, POSTECH, Pohang, Gyeongbuk 37673, Korea}
\affiliation{Asia Pacific Center for Theoretical Physics (APCTP), Pohang, Gyeongbuk 37673, Korea}

\date{\today}

\begin{abstract}
We derive an effective dual holographic Einstein-Maxwell theory, applying renormalization group transformations to interacting Dirac fermions in a recursive way. In particular, we show how both background metric tensor and U(1) gauge fields become dynamical to describe the renormalization group flows of coupling functions and order parameter fields of the corresponding quantum field theory through natural emergence of an extra-dimensional space, respectively. Finally, we propose a prescription on how to calculate correlation functions in a non-perturbative way, where quantum fluctuations of both metric and gauge fields are frozen to cause a classical field theory of the Einstein-Maxwell type in the large $N$ limit. Here, $N$ is the flavor degeneracy of Dirac fermions. This prescription turns out to coincide with that of the holographic duality conjecture.
\end{abstract}


\maketitle

\section{Introduction}

Quasiparticles are fundamental building blocks for the perturbative theoretical framework. Their existence is allowed by sufficient number of emergent symmetries and protected from decaying into bunch of quasiparticle-quasihole excitations as long as such emergent symmetries are preserved. However, strong correlations between quasiparticles can arise through renormalization group flows. These effective interactions may give rise to predominantly fast relaxation and result in local equilibrium through transferring their energies into each other and mixing good quantum numbers between them. As a result, only minimal number of symmetries are preserved and dynamics of these non-quasiparticle systems are described by remaining conserved currents such as charge (particle number), momentum, and energy \cite{Holographic_Liquid_Son_I,Holographic_Liquid_Son_II,Holographic_Liquid_Son_III, Holographic_Liquid_Son_IV}.

Recently, this effective hydrodynamics has been observed in the system of quasi two-dimensional Dirac fermions \cite{Hydro_Exp_I,Hydro_Exp_II,Hydro_Exp_III}. More concretely, experiments measured typical hydrodynamic behaviors such as negative longitudinal resistivity, violation of the Wiedemann-Franz law, and fluid viscosity, respectively, in the hydrodynamic regime. In spite of these clear features on effective hydrodynamics, we would like to point out that the physical origin has not been completely clarified. More precisely, it is not clear how the relaxation rate due to inelastic collisions between electrons prevails over either that between electrons and phonons or the elastic scattering rate between electrons and nonmagnetic impurities, responsible for such emergent hydrodynamic phenomena. One has to investigate the renormalization group flows of these three time scales in a self-consistent way. However, validity of the renormalization group analysis may be beyond the perturbative regime.

Besides possible phenomenological approaches for effective hydrodynamics, where transport coefficients are taken to be fitting parameters for experiments \cite{Kinetic_Theory_D_Tong}, it is natural to consider the dual holographic approach as a non-perturbative theoretical framework \cite{Holographic_Duality_I,Holographic_Duality_II,Holographic_Duality_III,Holographic_Duality_IV,Holographic_Duality_V,Holographic_Duality_VI, Holographic_Duality_VII}. The holographic dual description gives rise to effective hydrodynamic phenomena in $D$ spacetime dimensions, formulated as an effective classical field theory on an emergent curved spacetime in $D+1$ dimensions and regarded to be dual to strongly coupled quantum field theories in $D$ spacetime dimensions. For example \cite{Holographic_Liquid_KSK}, solving the Maxwell equation coupled to the Einstein equation, referred to as Einstein-Maxwell theory in the holographic duality conjecture, those classical solutions describe not only diffusive dynamics in transverse modes but also sound dynamics in longitudinal modes as those modes in the hydrodynamics \cite{Holographic_Liquid_Son_I,Holographic_Liquid_Son_II,Holographic_Liquid_Son_III, Holographic_Liquid_Son_IV}. In particular, the effective hydrodynamics in the holographic approach turns out to give a universally small value of the $\eta / s$ ratio, where $\eta$ is shear viscosity and $s$ is entropy \cite{Holographic_Liquid_Son_IV}. One may call this emergent fluid holographic liquid. This holographic-liquid description could explain the violation of the Wiedemann-Franz law in graphene rather remarkably, which fits both electrical and thermal transport coefficients near the charge-neutral point in the regime of intermediate temperatures based on two fitting parameters \cite{Holographic_Liquid_Graphene}.

In the present study, we propose how to derive the holographic-liquid dual description, more precisely, an effective Einstein-Maxwell theory from an effective quantum field theory of interacting Dirac fermions. We take Wilsonian renormalization group transformations \cite{RG_Textbook} recursively not in momentum space but in real space a la Polchinski \cite{RG_realspace_Polchinski}, further developed by Sung-Sik Lee in his emergent holographic construction \cite{SungSik_Holography_I,SungSik_Holography_II,SungSik_Holography_III}. In particular, we show that both background metric and U(1) electromagnetic gauge fields turn into dynamical quantum objects through their evolutions in an emergent extra-dimensional space, where the infinitesimal distance in the extra dimensional space is identified with an energy scale for renormalization group transformations. In other words, global symmetries are uplifted into gauge symmetries through natural emergence of this extra dimensional space. It turns out that quantum fluctuations of both metric and gauge fields are frozen to cause a classical field theory of the Einstein-Maxwell type in the large $N$ limit, where $N$ is the flavor degeneracy of Dirac fermions. Our derivation of the holographic dual effective field theory implies that effective Einstein equations describe the renormalization group flows of coupling functions along the extra dimension in the corresponding quantum field theory \cite{Holographic_Description_Kim,Holographic_Description_Kim_Ryu,RG_GR_Holography_Kim}. On the other hand, we point out that effective Maxwell equations describe the evolution of U(1) conserved currents along the extra dimension, which may be identified with the Callan-Symanzik equation of an order parameter field in the dual description. Finally, we propose a prescription on how to calculate correlation functions in a non-perturbative way \cite{Holographic_Liquid_Kim}, solving coupled differential equations of the Einstein-Maxwell type theory with boundary conditions. This prescription coincides with that of the dual holographic description \cite{Holographic_Duality_I,Holographic_Duality_II,Holographic_Duality_III, Holographic_Duality_IV,Holographic_Duality_V,Holographic_Duality_VI,Holographic_Duality_VII}.

\section{Effective geometric description for interacting Dirac fermions}

\subsection{Interacting Dirac fermions in a background metric tensor and an external U(1) electromagnetic gauge field}

We consider dynamics of interacting Dirac fermions in a background metric tensor and an external U(1) electromagnetic gauge field \cite{Coupling_Scalarfields_Riccicurvature}
\bqa && Z[\phi_{\alpha}, A_{B \mu}] = \int D \psi_{\alpha} \exp\Big[ - \int d^{D} x \sqrt{g_{B}} \Big\{ \bar{\psi}_{\alpha} \gamma^{a} e_{B a}^{\mu} \Big( \partial_{\mu} - i q A_{B \mu} - \frac{i}{4} \omega_{B \mu}^{a'b'} \sigma_{a'b'} \Big) \psi_{\alpha} + m \bar{\psi}_{\alpha} \psi_{\alpha} \nn && + \frac{\lambda_{\chi}}{2 N} (\bar{\psi}_{\alpha} \psi_{\alpha}) (\bar{\psi}_{\beta} \psi_{\beta}) + \frac{q \lambda_{j}}{2 N} g^{\mu\nu}_{B} (\bar{\psi}_{\alpha} \gamma_{a} e_{B \mu}^{a} \psi_{\alpha}) (\bar{\psi}_{\beta} \gamma_{b} e_{B \nu}^{b} \psi_{\beta}) + ( \bar{\phi}_{\alpha} p_{B}^{\dagger} \psi_{\alpha} + \bar{\psi}_{\alpha} p_{B} \phi_{\alpha}) \Big\} \Big] . \eqa
Here, $\psi_{\alpha}$ is a Dirac spinor at $x$ in $D$ spacetime dimensions. $\alpha$ runs from $1$ to $N$, denoting the flavor degeneracy of Dirac fermions. $\gamma^{a}$ is a Dirac $\gamma$ matrix, defined in a local rest frame at $x$ and satisfying the Clifford algebra $\{\gamma^{a}, \gamma^{b}\} = 2 \delta^{ab}$ with the Euclidean signature. $e_{B a}^{\mu}$ defines the local rest frame given by the tangent manifold at $x$, called vierbein. The corresponding background metric is given by the vierbein as follows $g_{B \mu\nu} = e_{B \mu}^{a} e_{B \nu}^{b} \delta_{ab}$. $A_{B \mu}$ is an external electromagnetic U(1) gauge field and $q$ is a unit electromagnetic charge of $\psi_{\alpha}$. $\omega_{B \mu}^{ab} = e_{B \nu}^{a} \partial_{\mu} e_{B}^{\nu b} + e_{B \nu}^{a} \Gamma^{\nu}_{B \sigma\mu} e_{B}^{\sigma b}$ is a background spin connection and $\sigma_{ab} = \frac{i}{2} [\gamma^{a}, \gamma^{b}]$ is a commutator of Dirac gamma matrixes in the local rest frame. Here, $\Gamma_{B \sigma\mu}^{\nu} = \frac{1}{2} g^{B \nu\delta} (\partial_{\sigma} g_{B \delta\mu} + \partial_{\mu} g_{B \sigma\delta} - \partial_{\delta} g_{B \sigma\mu})$ is the Christoffel symbol. $m$ represents a mass of Dirac fermions. $\lambda_{\chi}$ is the coupling constant of an effective interaction term for dynamical mass generation \cite{Dynamical_Mass_Generation_QED} and $\lambda_{j}$ is that of an effective interaction term between U(1) conserved currents. $\phi_{\alpha}$ has been introduced as a source field to help us calculate the Green's function of these interacting Dirac fermions, where $p_{B}$ is a matrix-valued coupling constant between the source field and the Dirac fermion field, which satisfies $\gamma^{0} p_{B}^{\dagger} \gamma^{0} = p_{B}^{\dagger}$ given by hermiticity.

%
%

%
%

Performing the Hubbard-Stratonovich transformation for both effective interactions, we reformulate the above expression as follows
\bqa && Z[\phi_{\alpha}, A_{B \mu}] = \int D \psi_{\alpha} D \varphi^{(0)} D a_{\mu}^{(0)} D e_{a}^{\mu (0)} D \theta^{a (0)}_{\mu} D p^{(0)} D q^{(0)} \nn && \exp\Big[ - \int d^{D} x \sqrt{g^{(0)}} \Big\{ \bar{\psi}_{\alpha} \gamma^{a} e_{a}^{\mu (0)} \Big( \partial_{\mu} - i q a_{\mu}^{(0)} - \frac{i}{4} \omega_{\mu}^{a'b' (0)} \sigma_{a'b'} \Big) \psi_{\alpha} + (m - i \varphi^{(0)}) \bar{\psi}_{\alpha} \psi_{\alpha} \nn && + ( \bar{\phi}_{\alpha} p^{(0) \dagger} \psi_{\alpha} + \bar{\psi}_{\alpha} p^{(0)} \phi_{\alpha}) + \frac{N}{2 \lambda_{\chi}} \varphi^{(0) 2} + \frac{N q}{2 \lambda_{j}} g^{\mu\nu (0)} (a_{\mu}^{(0)} - A_{B \mu}) (a_{\nu}^{(0)} - A_{B \nu}) \nn && - N \theta_{\mu}^{a (0)} (e_{a}^{\mu (0)} - e_{B a}^{\mu}) - N [q^{(0) \dagger} (p^{(0)} - p_{B}) + H.c.] \Big\} \Big] . \eqa
Here, $e_{a}^{\mu (0)}$ is a dynamically introduced vierbein field and its dual tensor field $\theta_{\mu}^{a (0)}$ is a Lagrange multiplier field to impose the constraint given by $\delta(e_{a}^{\mu (0)} - e_{B a}^{\mu})$. $\varphi^{(0)}$ is an order-parameter field dual to $\bar{\psi}_{\alpha} \psi_{\alpha}$ and $a_{\mu}^{(0)}$ is an effective U(1) gauge field dual to $\bar{\psi}_{\alpha} \gamma^{a} e_{a}^{\mu (0)} \psi_{\alpha}$. $q^{(0)}$ is a Lagrange multiplier field to impose the constraint given by $\delta(p^{(0)} - p_{B})$.

%
%

\subsection{Real space renormalization group transformation a la Polchinski}

\subsubsection{Renormalization group transformation for $\varphi^{(0)}$}

We perform a renormalization group transformation in real space a la Polchinski \cite{RG_realspace_Polchinski}, further developed by Sung-Sik Lee in his emergent holographic construction \cite{SungSik_Holography_I,SungSik_Holography_II,SungSik_Holography_III}. First, we consider the renormalization group transformation for the order parameter field $\varphi^{(0)}$. We introduce an auxiliary order-parameter field $\sigma^{(0)}$ with its mass $\frac{N}{2 \lambda_{\sigma}}$ as follows
\bqa && Z[\phi_{\alpha}, A_{B \mu}] = \int D \psi_{\alpha} D \varphi^{(0)} D \sigma^{(0)} D a_{\mu}^{(0)} D e_{a}^{\mu (0)} D \theta^{a (0)}_{\mu} D p^{(0)} D q^{(0)} \nn && \exp\Big[ - \int d^{D} x \sqrt{g^{(0)}} \Big\{ \bar{\psi}_{\alpha} \gamma^{a} e_{a}^{\mu (0)} \Big( \partial_{\mu} - i q a_{\mu}^{(0)} - \frac{i}{4} \omega_{\mu}^{a'b' (0)} \sigma_{a'b'} \Big) \psi_{\alpha} + (m - i \varphi^{(0)}) \bar{\psi}_{\alpha} \psi_{\alpha} \nn && + ( \bar{\phi}_{\alpha} p^{(0) \dagger} \psi_{\alpha} + \bar{\psi}_{\alpha} p^{(0)} \phi_{\alpha}) + \frac{N}{2 \lambda_{\chi}} \varphi^{(0) 2} + \frac{N}{2 \lambda_{\sigma}} \sigma^{(0) 2} + \frac{N q}{2 \lambda_{j}} g^{\mu\nu (0)} (a_{\mu}^{(0)} - A_{B \mu}) (a_{\nu}^{(0)} - A_{B \nu}) \nn && - N \theta_{\mu}^{a (0)} (e_{a}^{\mu (0)} - e_{B a}^{\mu}) - N [q^{(0) \dagger} (p^{(0)} - p_{B}) + H.c.] \Big\} \Big] . \eqa
We point out that the introduction of the auxiliary field $\sigma^{(0)}$ in this way does not alter the original physics at all except for the normalization constant in the partition function, here omitted for simplicity.

To perform the renormalization group transformation, we separate these two original fields into their slow and fast degrees of freedom in the following way
\bqa && \varphi^{(0)} \Longrightarrow \varphi^{(0)} + \Phi^{(0)} , ~~~~~ \sigma^{(0)} \Longrightarrow c_{\varphi}^{(0)} \varphi^{(0)} + c_{\Phi}^{(0)} \Phi^{(0)} . \eqa
Here, $\varphi^{(0)}$ and $\Phi^{(0)}$ correspond to slow and fast degrees of freedom, respectively, which will be clarified below. Two coefficients of $c_{\varphi}^{(0)}$ and $c_{\Phi}^{(0)}$ are determined by the fact that possible cross terms in the following expression
\bqa && \frac{N}{2 \lambda_{\chi}} \varphi^{(0) 2} + \frac{N}{2 \lambda_{\sigma}} \sigma^{(0) 2} \Longrightarrow \frac{N}{2} \Big( \frac{1}{\lambda_{\chi}} + \frac{c_{\varphi}^{(0) 2}}{\lambda_{\sigma}} \Big) \varphi^{(0) 2} + \frac{N}{2} \Big( \frac{1}{\lambda_{\chi}} + \frac{c_{\Phi}^{(0) 2}}{\lambda_{\sigma}} \Big) \Phi^{(0) 2} + N \Big( \frac{1}{\lambda_{\chi}} + \frac{c_{\varphi}^{(0)} c_{\Phi}^{(0)}}{\lambda_{\sigma}} \Big) \varphi^{(0)} \Phi^{(0)} \eqa
do not appear, given by
\bqa && \frac{1}{\lambda_{\chi}} + \frac{c_{\varphi}^{(0)} c_{\Phi}^{(0)}}{\lambda_{\sigma}} = 0 . \eqa
As a result, we obtain
\bqa && c_{\varphi}^{(0)} = \frac{\lambda_{\chi}^{-1}}{\mu_{\varphi}^{(0)} \lambda_{\sigma}^{-1/2}} , ~~~~~ c_{\Phi}^{(0)} = - \frac{\mu_{\varphi}^{(0)}}{\lambda_{\sigma}^{-1/2}} . \eqa
Here, $\mu_{\varphi}^{(0)}$ is the effective mass of high-energy quantum fluctuations, given by
\bqa && \mu_{\varphi}^{(0)} = \frac{\lambda_{\chi}^{-1/2}}{\sqrt{e^{2 \beta^{(0)} d z} - 1}} , \eqa
where $\beta^{(0)}$ represents the local speed of coarse graining, which corresponds to the lapse function in the so called ADM Hamiltonian-gravity formulation \cite{ADM_Hamiltonian_Gravity}. $d z$ is an infinitesimal parameter, which controls the path integral for heavy order-parameter fields. The resulting mass terms are
\bqa && \frac{N}{2 \lambda_{\chi}} \varphi^{(0) 2} + \frac{N}{2 \lambda_{\sigma}} \sigma^{(0) 2} = \frac{N}{2 \lambda_{\chi}} e^{2 \beta^{(0)} d z} \varphi^{(0) 2} + \frac{N}{2 \lambda_{\chi}} \frac{e^{2 \beta^{(0)} d z}}{e^{2 \beta^{(0)} d z} - 1} \Phi^{(0) 2} , \eqa
which identifies $\varphi^{(0)}$ and $\Phi^{(0)}$ with low-energy and high-energy quantum fluctuations, respectively. Rescaling both order-parameter fields as
\bqa && \varphi^{(0)} \Longrightarrow e^{- \beta^{(0)} d z} \varphi^{(0)} , ~~~~~ \Phi^{(0)} \Longrightarrow e^{- \beta^{(0)} d z} \Phi^{(0)} , \eqa
which returns the mass term of low-energy order-parameter fluctuations to its original expression, we obtain
\bqa && Z[\phi_{\alpha}, A_{B \mu}] = \int D \psi_{\alpha} D \varphi^{(0)} D \Phi^{(0)} D a_{\mu}^{(0)} D e_{a}^{\mu (0)} D \theta^{a (0)}_{\mu} D p^{(0)} D q^{(0)} \nn && \exp\Big[ - \int d^{D} x \sqrt{g^{(0)}} \Big\{ \bar{\psi}_{\alpha} \gamma^{a} e_{a}^{\mu (0)} \Big( \partial_{\mu} - i q a_{\mu}^{(0)} - \frac{i}{4} \omega_{\mu}^{a'b' (0)} \sigma_{a'b'} \Big) \psi_{\alpha} + (m - i e^{- \beta^{(0)} d z} \varphi^{(0)} - i e^{- \beta^{(0)} d z} \Phi^{(0)}) \bar{\psi}_{\alpha} \psi_{\alpha} \nn && + ( \bar{\phi}_{\alpha} p^{(0) \dagger} \psi_{\alpha} + \bar{\psi}_{\alpha} p^{(0)} \phi_{\alpha}) + \frac{N}{2 \lambda_{\chi}} \varphi^{(0) 2} + \frac{N}{2 \lambda_{\chi}} \frac{1}{e^{2 \beta^{(0)} d z} - 1} \Phi^{(0) 2} + \frac{N q}{2 \lambda_{j}} g^{\mu\nu (0)} (a_{\mu}^{(0)} - A_{B \mu}) (a_{\nu}^{(0)} - A_{B \nu}) \nn && - N \theta_{\mu}^{a (0)} (e_{a}^{\mu (0)} - e_{B a}^{\mu}) - N [q^{(0) \dagger} (p^{(0)} - p_{B}) + H.c.] \Big\} \Big] . \eqa

Performing the path integral $\int D \Phi^{(0)}$, we do the renormalization group transformation for the order-parameter field $\varphi^{(0)}$ and find
\bqa && Z[\phi_{\alpha}, A_{B \mu}] = \int D \psi_{\alpha} D \varphi^{(0)} D a_{\mu}^{(0)} D e_{a}^{\mu (0)} D \theta^{a (0)}_{\mu} D p^{(0)} D q^{(0)} \exp\Big[ - \frac{1}{2} \mbox{tr}_{xx'} \ln \frac{N}{2 \lambda_{\chi}} \frac{\sqrt{g^{(0)}}}{e^{2 \beta^{(0)} d z} - 1} \nn && - \int d^{D} x \sqrt{g^{(0)}} \Big\{ \bar{\psi}_{\alpha} \gamma^{a} e_{a}^{\mu (0)} \Big( \partial_{\mu} - i q a_{\mu}^{(0)} - \frac{i}{4} \omega_{\mu}^{a'b' (0)} \sigma_{a'b'} \Big) \psi_{\alpha} + (m - i e^{- \beta^{(0)} d z} \varphi^{(0)}) \bar{\psi}_{\alpha} \psi_{\alpha} \nn && + ( \bar{\phi}_{\alpha} p^{(0) \dagger} \psi_{\alpha} + \bar{\psi}_{\alpha} p^{(0)} \phi_{\alpha}) + \frac{N}{2 \lambda_{\chi}} \varphi^{(0) 2} + \frac{\lambda_{\chi}}{2 N} (1 - e^{- 2 \beta^{(0)} d z}) (\bar{\psi}_{\alpha} \psi_{\alpha}) (\bar{\psi}_{\beta} \psi_{\beta}) + \frac{N q}{2 \lambda_{j}} g^{\mu\nu (0)} (a_{\mu}^{(0)} - A_{B \mu}) (a_{\nu}^{(0)} - A_{B \nu}) \nn && - N \theta_{\mu}^{a (0)} (e_{a}^{\mu (0)} - e_{B a}^{\mu}) - N [q^{(0) \dagger} (p^{(0)} - p_{B}) + H.c.] \Big\} \Big] , \eqa
where effective interactions $\frac{\lambda_{\chi}}{2 N} (1 - e^{- 2 \beta^{(0)} d z}) (\bar{\psi}_{\alpha} \psi_{\alpha}) (\bar{\psi}_{\beta} \psi_{\beta})$ are newly generated after the renormalization group transformation for $\varphi^{(0)}$. Here, an idea is to take into account these effective interactions in a recursive way. We perform the Hubbard-Stratonovich transformation for this effective interaction term and obtain the following recursive expression of the partition function for the order-parameter field
\bqa && Z[\phi_{\alpha}, A_{B \mu}] = \int D \psi_{\alpha} D \varphi^{(0)} D \varphi^{(1)} D a_{\mu}^{(0)} D e_{a}^{\mu (0)} D \theta^{a (0)}_{\mu} D p^{(0)} D q^{(0)} \exp\Big[ - \frac{1}{2} \mbox{tr}_{xx'} \ln \frac{N}{2 \lambda_{\chi}} \frac{\sqrt{g^{(0)}}}{e^{2 \beta^{(0)} d z} - 1} \nn && - \int d^{D} x \sqrt{g^{(0)}} \Big\{ \bar{\psi}_{\alpha} \gamma^{a} e_{a}^{\mu (0)} \Big( \partial_{\mu} - i q a_{\mu}^{(0)} - \frac{i}{4} \omega_{\mu}^{a'b' (0)} \sigma_{a'b'} \Big) \psi_{\alpha} + (m - i e^{- \beta^{(0)} d z} \varphi^{(1)}) \bar{\psi}_{\alpha} \psi_{\alpha} \nn && + ( \bar{\phi}_{\alpha} p^{(0) \dagger} \psi_{\alpha} + \bar{\psi}_{\alpha} p^{(0)} \phi_{\alpha}) + \frac{N}{2 \lambda_{\chi}} \varphi^{(0) 2} + \frac{N}{2 \lambda_{\chi}} \frac{1}{e^{2 \beta^{(0)} d z} - 1} (\varphi^{(1)} - \varphi^{(0)})^{2} + \frac{N q}{2 \lambda_{j}} g^{\mu\nu (0)} (a_{\mu}^{(0)} - A_{B \mu}) (a_{\nu}^{(0)} - A_{B \nu}) \nn && - N \theta_{\mu}^{a (0)} (e_{a}^{\mu (0)} - e_{B a}^{\mu}) - N [q^{(0) \dagger} (p^{(0)} - p_{B}) + H.c.] \Big\} \Big] , \eqa
where the order-parameter field $\varphi^{(1)}$ has been shifted as
\bqa && \varphi^{(1)} \Longrightarrow \varphi^{(1)} - \varphi^{(0)} . \eqa
The recursive expression means that repeating the renormalization group transformation for $\varphi^{(1)}$ instead of $\varphi^{(0)}$ as the second iteration procedure gives rise to the recursive formula for the effective action, which will be clarified below.

\subsubsection{Renormalization group transformation for $a_{\mu}^{(0)}$}

Second, we perform the renormalization group transformation for quantum fluctuations of U(1) gauge fields. We introduce an auxiliary U(1) gauge field $A_{\mu}^{(0)}$ in the following way
\bqa && Z[\phi_{\alpha}, A_{B \mu}] = \int D \psi_{\alpha} D \varphi^{(0)} D \varphi^{(1)} D a_{\mu}^{(0)} D A_{\mu}^{(0)} D e_{a}^{\mu (0)} D \theta^{a (0)}_{\mu} D p^{(0)} D q^{(0)} \exp\Big[ - \frac{1}{2} \mbox{tr}_{xx'} \ln \frac{N}{2 \lambda_{\chi}} \frac{\sqrt{g^{(0)}}}{e^{2 \beta^{(0)} d z} - 1} \nn && - \int d^{D} x \sqrt{g^{(0)}} \Big\{ \bar{\psi}_{\alpha} \gamma^{a} e_{a}^{\mu (0)} \Big( \partial_{\mu} - i q [a_{\mu}^{(0)} + A_{B \mu}] - \frac{i}{4} \omega_{\mu}^{a'b' (0)} \sigma_{a'b'} \Big) \psi_{\alpha} + (m - i e^{- \beta^{(0)} d z} \varphi^{(1)}) \bar{\psi}_{\alpha} \psi_{\alpha} \nn && + ( \bar{\phi}_{\alpha} p^{(0) \dagger} \psi_{\alpha} + \bar{\psi}_{\alpha} p^{(0)} \phi_{\alpha}) + \frac{N}{2 \lambda_{\chi}} \varphi^{(0) 2} + \frac{N}{2 \lambda_{\chi}} \frac{1}{e^{2 \beta^{(0)} d z} - 1} (\varphi^{(1)} - \varphi^{(0)})^{2} + \frac{N q}{2 \lambda_{j}} g^{\mu\nu (0)} a_{\mu}^{(0)} a_{\nu}^{(0)} + \frac{N q}{2 \lambda_{A}} g^{\mu\nu (0)} A_{\mu}^{(0)} A_{\nu}^{(0)} \nn && - N \theta_{\mu}^{a (0)} (e_{a}^{\mu (0)} - e_{B a}^{\mu}) - N [q^{(0) \dagger} (p^{(0)} - p_{B}) + H.c.] \Big\} \Big] , \eqa
where $a_{\mu}^{(0)} \Longrightarrow a_{\mu}^{(0)} + A_{B \mu}$ has been done. Again, we also mention that this introduction of auxiliary U(1) gauge fields does not change the partition function except for the normalization constant, omitted for simplicity.

Separating two U(1) gauge fields into their slow and fast degrees of freedom in the following way
\bqa && a_{\mu}^{(0)} \Longrightarrow a_{\mu}^{(0)} + \mathcal{A}_{\mu}^{(0)} , ~~~~~ A_{\mu}^{(0)} \Longrightarrow c_{a}^{(0)} a_{\mu}^{(0)} + c_{\mathcal{A}}^{(0)} \mathcal{A}_{\mu}^{(0)} , \eqa
we rewrite the two Gaussian-fluctuation terms of U(1) gauge fields as
\bqa && \frac{N q}{2 \lambda_{j}} g^{\mu\nu (0)} a_{\mu}^{(0)} a_{\nu}^{(0)} + \frac{N q}{2 \lambda_{A}} g^{\mu\nu (0)} A_{\mu}^{(0)} A_{\nu}^{(0)} = \frac{N q}{2 \lambda_{j}} e^{2 \delta^{(0)} d z} g^{\mu\nu (0)} a_{\mu}^{(0)} a_{\nu}^{(0)} + \frac{N q}{2 \lambda_{j}} \frac{e^{2 \delta^{(0)} d z}}{e^{2 \delta^{(0)} d z} - 1} g^{\mu\nu (0)} \mathcal{A}_{\mu}^{(0)} \mathcal{A}_{\nu}^{(0)} , \eqa
where two coefficients are given by
\bqa && c_{a}^{(0)} = \frac{\lambda_{j}^{-1}}{\mu_{a}^{(0)} \lambda_{A}^{-1/2}} , ~~~~~ c_{\mathcal{A}}^{(0)} = - \frac{\mu_{a}^{(0)}}{\lambda_{A}^{-1/2}} , \eqa
which makes possible cross terms vanish as shown in the previous section. $\mu_{a}^{(0)}$ is an effective mass parameter, given by
\bqa && \mu_{a}^{(0)} = \frac{\lambda_{j}^{-1/2}}{\sqrt{e^{2 \delta^{(0)} d z} - 1}} \eqa
and introduced for the renormalization group transformation. $\delta^{(0)}$ is the local speed of coarse graining with an infinitesimal parameter $d z$ for the renormalization group transformation in U(1) gauge fluctuations. Rescaling both low-energy and high-energy quantum fluctuations of U(1) gauge fields as
\bqa && a_{\mu}^{(0)} \Longrightarrow e^{- \delta^{(0)} d z} a_{\mu}^{(0)} , ~~~~~ \mathcal{A}_{\mu}^{(0)} \Longrightarrow e^{- \delta^{(0)} d z} \mathcal{A}_{\mu}^{(0)} , \eqa
which returns the mass term of slow fluctuations of U(1) gauge fields to its original expression, we obtain
\bqa && Z[\phi_{\alpha}, A_{B \mu}] = \int D \psi_{\alpha} D \varphi^{(0)} D \varphi^{(1)} D a_{\mu}^{(0)} D \mathcal{A}_{\mu}^{(0)} D e_{a}^{\mu (0)} D \theta^{a (0)}_{\mu} D p^{(0)} D q^{(0)} \exp\Big[ - \frac{1}{2} \mbox{tr}_{xx'} \ln \frac{N}{2 \lambda_{\chi}} \frac{\sqrt{g^{(0)}}}{e^{2 \beta^{(0)} d z} - 1} \nn && - \int d^{D} x \sqrt{g^{(0)}} \Big\{ \bar{\psi}_{\alpha} \gamma^{a} e_{a}^{\mu (0)} \Big( \partial_{\mu} - i q [e^{- \delta^{(0)} d z} a_{\mu}^{(0)} + A_{B \mu} + e^{- \delta^{(0)} d z} \mathcal{A}_{\mu}^{(0)}] - \frac{i}{4} \omega_{\mu}^{a'b' (0)} \sigma_{a'b'} \Big) \psi_{\alpha} \nn && + (m - i e^{- \beta^{(0)} d z} \varphi^{(1)}) \bar{\psi}_{\alpha} \psi_{\alpha} + ( \bar{\phi}_{\alpha} p^{(0) \dagger} \psi_{\alpha} + \bar{\psi}_{\alpha} p^{(0)} \phi_{\alpha}) + \frac{N}{2 \lambda_{\chi}} \varphi^{(0) 2} + \frac{N}{2 \lambda_{\chi}} \frac{1}{e^{2 \beta^{(0)} d z} - 1} (\varphi^{(1)} - \varphi^{(0)})^{2} \nn && + \frac{N q}{2 \lambda_{j}} g^{\mu\nu (0)} a_{\mu}^{(0)} a_{\nu}^{(0)} + \frac{N q}{2 \lambda_{j}} \frac{1}{e^{2 \delta^{(0)} d z} - 1} g^{\mu\nu (0)} \mathcal{A}_{\mu}^{(0)} \mathcal{A}_{\nu}^{(0)} - N \theta_{\mu}^{a (0)} (e_{a}^{\mu (0)} - e_{B a}^{\mu}) - N [q^{(0) \dagger} (p^{(0)} - p_{B}) + H.c.] \Big\} \Big] . \nn \eqa

It is straightforward to perform the Gaussian integral for such high-energy quantum fluctuations of U(1) gauge fields, given by
\bqa && Z[\phi_{\alpha}, A_{B \mu}] = \int D \psi_{\alpha} D \varphi^{(0)} D \varphi^{(1)} D a_{\mu}^{(0)} D e_{a}^{\mu (0)} D \theta^{a (0)}_{\mu} D p^{(0)} D q^{(0)} \nn && \exp\Big[ - \frac{1}{2} \mbox{tr}_{xx'} \ln \frac{N}{2 \lambda_{\chi}} \frac{\sqrt{g^{(0)}}}{e^{2 \beta^{(0)} d z} - 1} - \frac{1}{2} \mbox{tr}_{xx'} \ln \frac{N q}{2 \lambda_{j}} \frac{\sqrt{g^{(0)}}}{e^{2 \delta^{(0)} d z} - 1} g^{\mu\nu (0)} \nn && - \int d^{D} x \sqrt{g^{(0)}} \Big\{ \bar{\psi}_{\alpha} \gamma^{a} e_{a}^{\mu (0)} \Big( \partial_{\mu} - i q [e^{- \delta^{(0)} d z} a_{\mu}^{(0)} + A_{B \mu}] - \frac{i}{4} \omega_{\mu}^{a'b' (0)} \sigma_{a'b'} \Big) \psi_{\alpha} + (m - i e^{- \beta^{(0)} d z} \varphi^{(1)}) \bar{\psi}_{\alpha} \psi_{\alpha} \nn && + ( \bar{\phi}_{\alpha} p^{(0) \dagger} \psi_{\alpha} + \bar{\psi}_{\alpha} p^{(0)} \phi_{\alpha}) + \frac{N}{2 \lambda_{\chi}} \varphi^{(0) 2} + \frac{N}{2 \lambda_{\chi}} \frac{1}{e^{2 \beta^{(0)} d z} - 1} (\varphi^{(1)} - \varphi^{(0)})^{2} + \frac{N q}{2 \lambda_{j}} g^{\mu\nu (0)} a_{\mu}^{(0)} a_{\nu}^{(0)} \nn && + \frac{q \lambda_{j}}{2 N} (1 - e^{- 2 \delta^{(0)} d z}) g^{\mu\nu (0)} (\bar{\psi}_{\alpha} \gamma_{a} e_{\mu}^{a (0)} \psi_{\alpha}) (\bar{\psi}_{\beta} \gamma_{b} e_{\nu}^{b (0)} \psi_{\beta}) - N \theta_{\mu}^{a (0)} (e_{a}^{\mu (0)} - e_{B a}^{\mu}) - N [q^{(0) \dagger} (p^{(0)} - p_{B}) + H.c.] \Big\} \Big] . \nn \eqa
Again, effective current-current interactions $\frac{q \lambda_{j}}{2 N} (1 - e^{- 2 \delta^{(0)} d z}) g^{\mu\nu (0)} (\bar{\psi}_{\alpha} \gamma_{a} e_{\mu}^{a (0)} \psi_{\alpha}) (\bar{\psi}_{\beta} \gamma_{b} e_{\nu}^{b (0)} \psi_{\beta})$ are also newly generated. Performing the Hubbard-Stratonovich transformation for this effective current-interaction term, we obtain the following recursive expression of the partition function
\bqa && Z[\phi_{\alpha}, A_{B \mu}] = \int D \psi_{\alpha} D \varphi^{(0)} D \varphi^{(1)} D a_{\mu}^{(0)} D \tilde{a}_{\mu}^{(1)} D e_{a}^{\mu (0)} D \theta^{a (0)}_{\mu} D p^{(0)} D q^{(0)} \nn && \exp\Big[ - \frac{1}{2} \mbox{tr}_{xx'} \ln \frac{N}{2 \lambda_{\chi}} \frac{\sqrt{g^{(0)}}}{e^{2 \beta^{(0)} d z} - 1} - \frac{1}{2} \mbox{tr}_{xx'} \ln \frac{N q}{2 \lambda_{j}} \frac{\sqrt{g^{(0)}}}{e^{2 \delta^{(0)} d z} - 1} g^{\mu\nu (0)} \nn && - \int d^{D} x \sqrt{g^{(0)}} \Big\{ \bar{\psi}_{\alpha} \gamma^{a} e_{a}^{\mu (0)} \Big( \partial_{\mu} - i q e^{- \delta^{(0)} d z} \tilde{a}_{\mu}^{(1)} - \frac{i}{4} \omega_{\mu}^{a'b' (0)} \sigma_{a'b'} \Big) \psi_{\alpha} + (m - i e^{- \beta^{(0)} d z} \varphi^{(1)}) \bar{\psi}_{\alpha} \psi_{\alpha} \nn && + ( \bar{\phi}_{\alpha} p^{(0) \dagger} \psi_{\alpha} + \bar{\psi}_{\alpha} p^{(0)} \phi_{\alpha}) + \frac{N}{2 \lambda_{\chi}} \varphi^{(0) 2} + \frac{N}{2 \lambda_{\chi}} \frac{1}{e^{2 \beta^{(0)} d z} - 1} (\varphi^{(1)} - \varphi^{(0)})^{2} \nn && + \frac{N q}{2 \lambda_{j}} g^{\mu\nu (0)} (a_{\mu}^{(0)} - e^{\delta^{(0)} d z} A_{B \mu}) (a_{\nu}^{(0)} - e^{\delta^{(0)} d z} A_{B \nu}) + \frac{N q}{2 \lambda_{j}} \frac{1}{e^{2 \delta^{(0)} d z} - 1} g^{\mu\nu (0)} (\tilde{a}_{\mu}^{(1)} - a_{\mu}^{(0)}) (\tilde{a}_{\nu}^{(1)} - a_{\nu}^{(0)}) \nn && - N \theta_{\mu}^{a (0)} (e_{a}^{\mu (0)} - e_{B a}^{\mu}) - N [q^{(0) \dagger} (p^{(0)} - p_{B}) + H.c.] \Big\} \Big] , \eqa
where
\bqa && \tilde{a}_{\mu}^{(1)} \Longrightarrow \tilde{a}_{\mu}^{(1)} - a_{\mu}^{(0)} \eqa
has been taken into account together with $a_{\mu}^{(0)} \Longrightarrow a_{\mu}^{(0)} - e^{\delta^{(0)} d z} A_{B \mu}$. The second iteration step for U(1) gauge fluctuations is to perform the renormalization group transformation for $\tilde{a}_{\mu}^{(1)}$ instead of that for $a_{\mu}^{(0)}$.

\subsubsection{Renormalization group transformation for $\psi_{\alpha}$}

Third, we perform the renormalization group transformation for Dirac fermions. We introduce an auxiliary Dirac spinor $\Phi_{\alpha}$ with its mass $M$ as follows
\bqa && Z[\phi_{\alpha}, A_{B \mu}] = \int D \psi_{\alpha} D \Phi_{\alpha} D \varphi^{(0)} D \varphi^{(1)} D a_{\mu}^{(0)} D \tilde{a}_{\mu}^{(1)} D e_{a}^{\mu (0)} D \theta^{a (0)}_{\mu} D p^{(0)} D q^{(0)} \nn && \exp\Big[ - \int d^{D} x \sqrt{g^{(0)}} \Big\{ \bar{\psi}_{\alpha} \gamma^{a} e_{a}^{\mu (0)} \Big( \partial_{\mu} - i q e^{- \delta^{(0)} d z} \tilde{a}_{\mu}^{(1)} - \frac{i}{4} \omega_{\mu}^{a'b' (0)} \sigma_{a'b'} \Big) \psi_{\alpha} + (m - i e^{- \beta^{(0)} d z} \varphi^{(1)}) \bar{\psi}_{\alpha} \psi_{\alpha} + M \bar{\Phi}_{\alpha} \Phi_{\alpha} \nn && + ( \bar{\phi}_{\alpha} p^{(0) \dagger} \psi_{\alpha} + \bar{\psi}_{\alpha} p^{(0)} \phi_{\alpha}) + \frac{N}{2 \lambda_{\chi}} \varphi^{(0) 2} + \frac{N}{2 \lambda_{\chi}} \frac{1}{e^{2 \beta^{(0)} d z} - 1} (\varphi^{(1)} - \varphi^{(0)})^{2} \nn && + \frac{N q}{2 \lambda_{j}} g^{\mu\nu (0)} (a_{\mu}^{(0)} - e^{\delta^{(0)} d z} A_{B \mu}) (a_{\nu}^{(0)} - e^{\delta^{(0)} d z} A_{B \nu}) + \frac{N q}{2 \lambda_{j}} \frac{1}{e^{2 \delta^{(0)} d z} - 1} g^{\mu\nu (0)} (\tilde{a}_{\mu}^{(1)} - a_{\mu}^{(0)}) (\tilde{a}_{\nu}^{(1)} - a_{\nu}^{(0)}) \nn && - N \theta_{\mu}^{a (0)} (e_{a}^{\mu (0)} - e_{B a}^{\mu}) - N [q^{(0) \dagger} (p^{(0)} - p_{B}) + H.c.] \Big\} \Big] , \eqa
where sub-leading constant terms of $\frac{1}{2} \mbox{tr}_{xx'} \ln \frac{N}{2 \lambda_{\chi}} \frac{\sqrt{g^{(0)}}}{e^{2 \beta^{(0)} d z} - 1} + \frac{1}{2} \mbox{tr}_{xx'} \ln \frac{N q}{2 \lambda_{j}} \frac{\sqrt{g^{(0)}}}{e^{2 \delta^{(0)} d z} - 1} g^{\mu\nu (0)}$ in the previous partition function have been neglected in the large $N$ limit. We emphasize that the introduction of the auxiliary Dirac spinor in this way does not alter the original physics at all except for the normalization constant in the partition function, here omitted for simplicity.

Now, we decompose these Dirac spinor fields in the following way
\bqa && \psi_{\alpha} \Longrightarrow \psi_{\alpha} + \Psi_{\alpha} , ~~~~~ \Phi_{\alpha} \Longrightarrow c_{\psi}^{(0)} \psi_{\alpha} + c_{\Psi}^{(0)} \Psi_{\alpha} \eqa
$\psi_{\alpha}$ is a Dirac spinor field with a light mass and $\Psi_{\alpha}$ is that with a heavy mass. Two positive coefficients of $c_{\psi}^{(0)}$ and $c_{\Psi}^{(0)}$ are determined by the fact that all mixing terms between $\psi_{\alpha}$ and $\Psi_{\alpha}$ do not appear in the two mass terms, given by
\bqa && (m - i e^{- \beta^{(0)} d z} \varphi^{(1)}) \bar{\psi}_{\alpha} \psi_{\alpha} + M \bar{\Phi}_{\alpha} \Phi_{\alpha} \nn && \Longrightarrow (m - i e^{- \beta^{(0)} d z} \varphi^{(1)} + M c_{\psi}^{(0) 2}) \bar{\psi}_{\alpha} \psi_{\alpha} + (m - i e^{- \beta^{(0)} d z} \varphi^{(1)} + M c_{\Psi}^{(0) 2}) \bar{\Phi}_{\alpha} \Phi_{\alpha} \nn && + (m - i e^{- \beta^{(0)} d z} \varphi^{(1)} + M c_{\psi}^{(0)} c_{\Psi}^{(0)}) (\bar{\psi}_{\alpha} \Psi_{\alpha} + \bar{\Psi}_{\alpha} \psi_{\alpha}) . \eqa
In other words, $m - i e^{- \beta^{(0)} d z} \varphi^{(1)} + M c_{\psi}^{(0)} c_{\Psi}^{(0)} = 0$ gives rise to
\bqa && c_{\psi}^{(0)} = \frac{m - i e^{- \beta^{(0)} d z} \varphi^{(1)}}{\mu_{\psi}^{(0) 1/2} M^{1/2}} , ~~~~~ c_{\Psi}^{(0)} = - \frac{\mu_{\psi}^{(0) 1/2}}{M^{1/2}} . \eqa
Here, the heavy mass $\mu_{\psi}^{(0)}$ is given by
\bqa && \mu_{\psi}^{(0) 1/2} = \frac{(m - i e^{- \beta^{(0)} d z} \varphi^{(1)})^{1/2}}{\sqrt{e^{2 \alpha^{(0)} d z} - 1}} , \eqa
where $d z$ is an infinitesimal parameter and $\alpha^{(0)}$ is the local speed of coarse graining, both of which control the path integral for heavy Dirac spinor fields. Then, we rewrite the mass terms with light and heavy Dirac spinor fields as
\bqa && (m - i e^{- \beta^{(0)} d z} \varphi^{(1)}) \bar{\psi}_{\alpha} \psi_{\alpha} + M \bar{\Phi}_{\alpha} \Phi_{\alpha} \nn && = e^{2 \alpha^{(0)} d z} (m - i e^{- \beta^{(0)} d z} \varphi^{(1)}) \bar{\psi}_{\alpha} \psi_{\alpha} + \frac{e^{2 \alpha^{(0)} d z}}{e^{2 \alpha^{(0)} d z} - 1} (m - i e^{- \beta^{(0)} d z} \varphi^{(1)}) \bar{\Psi}_{\alpha} \Psi_{\alpha} . \eqa

Taking rescaling of both Dirac spinor fields as
\bqa && \psi_{\alpha} \Longrightarrow e^{- \alpha^{(0)} d z} \psi_{\alpha} , ~~~~~ \Psi_{\alpha} \Longrightarrow e^{- \alpha^{(0)} d z} \Psi_{\alpha} , \eqa
which returns the mass term of light Dirac spinor fields into the original expression, we finish our setup for the renormalization group transformation of $\psi_{\alpha}$, given by
\bqa && Z[\phi_{\alpha}, A_{B \mu}] = \int D \psi_{\alpha} D \Psi_{\alpha} D \varphi^{(0)} D \varphi^{(1)} D a_{\mu}^{(0)} D \tilde{a}_{\mu}^{(1)} D e_{a}^{\mu (0)} D \theta^{a (0)}_{\mu} D p^{(0)} D q^{(0)} \nn && \exp\Big[ - \int d^{D} x \sqrt{g^{(0)}} \Big\{ e^{- 2 \alpha^{(0)} d z} \bar{\psi}_{\alpha} \gamma^{a} e_{a}^{\mu (0)} \Big( \partial_{\mu} - i q e^{- \delta^{(0)} d z} \tilde{a}_{\mu}^{(1)} - \frac{i}{4} \omega_{\mu}^{a'b' (0)} \sigma_{a'b'} \Big) \psi_{\alpha} \nn && + (m - i e^{- \beta^{(0)} d z} \varphi^{(1)}) \bar{\psi}_{\alpha} \psi_{\alpha} + e^{- \alpha^{(0)} d z} ( \bar{\phi}_{\alpha} p^{(0) \dagger} \psi_{\alpha} + \bar{\psi}_{\alpha} p^{(0)} \phi_{\alpha}) \nn && + e^{- 2 \alpha^{(0)} d z} \bar{\Psi}_{\alpha} \gamma^{a} e_{a}^{\mu (0)} \Big( \partial_{\mu} - i q e^{- \delta^{(0)} d z} \tilde{a}_{\mu}^{(1)} - \frac{i}{4} \omega_{\mu}^{a'b' (0)} \sigma_{a'b'} \Big) \Psi_{\alpha} + \frac{(m - i e^{- \beta^{(0)} d z} \varphi^{(1)})}{e^{2 \alpha^{(0)} d z} - 1} \bar{\Psi}_{\alpha} \Psi_{\alpha} \nn && + e^{- 2 \alpha^{(0)} d z} \Big\{ \bar{\psi}_{\alpha} \gamma^{a} e_{a}^{\mu (0)} \Big( \partial_{\mu} - i q e^{- \delta^{(0)} d z} \tilde{a}_{\mu}^{(1)} - \frac{i}{4} \omega_{\mu}^{a'b' (0)} \sigma_{a'b'} \Big) + e^{\alpha^{(0)} d z} \bar{\phi}_{\alpha} p^{(0) \dagger} \Big\} \Psi_{\alpha} \nn && + e^{- 2 \alpha^{(0)} d z} \bar{\Psi}_{\alpha} \Big\{ \gamma^{a} e_{a}^{\mu (0)} \Big( \partial_{\mu} - i q e^{- \delta^{(0)} d z} \tilde{a}_{\mu}^{(1)} - \frac{i}{4} \omega_{\mu}^{a'b' (0)} \sigma_{a'b'} \Big) \psi_{\alpha} + e^{\alpha^{(0)} d z} p^{(0)} \phi_{\alpha} \Big\} \nn && + \frac{N}{2 \lambda_{\chi}} \varphi^{(0) 2} + \frac{N}{2 \lambda_{\chi}} \frac{1}{e^{2 \beta^{(0)} d z} - 1} (\varphi^{(1)} - \varphi^{(0)})^{2} \nn && + \frac{N q}{2 \lambda_{j}} g^{\mu\nu (0)} \Big(a_{\mu}^{(0)} + \frac{i}{q} e^{\delta^{(0)} d z} d z \partial_{\mu} \alpha^{(0)} - e^{\delta^{(0)} d z} A_{B \mu}\Big) \Big(a_{\nu}^{(0)} + \frac{i}{q} e^{\delta^{(0)} d z} d z \partial_{\nu} \alpha^{(0)} - e^{\delta^{(0)} d z} A_{B \nu}\Big) \nn && + \frac{N q}{2 \lambda_{j}} \frac{1}{e^{2 \delta^{(0)} d z} - 1} g^{\mu\nu (0)} (\tilde{a}_{\mu}^{(1)} - a_{\mu}^{(0)}) (\tilde{a}_{\nu}^{(1)} - a_{\nu}^{(0)}) - N \theta_{\mu}^{a (0)} (e_{a}^{\mu (0)} - e_{B a}^{\mu}) - N [q^{(0) \dagger} (p^{(0)} - p_{B}) + H.c.] \Big\} \Big] . \eqa
Here, we introduced the gauge transformation as
\bqa && a_{\mu}^{(0)} \Longrightarrow a_{\mu}^{(0)} + \frac{i}{q} e^{\delta^{(0)} d z} d z \partial_{\mu} \alpha^{(0)} \eqa
into the above partition function.

%
%

It is also straightforward to perform the Gaussian path integral $\int D \Psi_{\alpha}$ for high-energy quantum fluctuations of Dirac spinor fields. As a result, we obtain
\bqa && Z[\phi_{\alpha x}, A_{B \mu x}] = \int D \psi_{\alpha x} D \varphi_{x}^{(0)} D \varphi_{x}^{(1)} D a_{\mu x}^{(0)} D \tilde{a}_{\mu x}^{(1)} D e_{a x}^{\mu (0)} D \theta^{a (0)}_{\mu x} D p_{x}^{(0)} D q_{x}^{(0)} \nn && \exp\Bigg[ N \mbox{tr}_{xx'} \ln \sqrt{g_{x}^{(0)}} e^{- 2 \alpha_{x}^{(0)} d z} \Big\{ \gamma^{a} e_{a x}^{\mu (0)} \Big( \partial_{\mu} - i q e^{- \delta_{x}^{(0)} d z} \tilde{a}_{\mu x}^{(1)} - \frac{i}{4} \omega_{\mu x}^{a'b' (0)} \sigma_{a'b'} \Big) + \frac{e^{2 \alpha_{x}^{(0)} d z} }{e^{2 \alpha_{x}^{(0)} d z} - 1} (m - i e^{- \beta_{x}^{(0)} d z} \varphi_{x}^{(1)}) \Big\} \nn && - \int d^{D} x \sqrt{g_{x}^{(0)}} \Bigg\{ e^{- 2 \alpha_{x}^{(0)} d z} \bar{\psi}_{\alpha x} \gamma^{a} e_{a x}^{\mu (0)} \Big( \partial_{\mu} - i q e^{- \delta_{x}^{(0)} d z} \tilde{a}_{\mu x}^{(1)} - \frac{i}{4} \omega_{\mu x}^{a'b' (0)} \sigma_{a'b'} \Big) \psi_{\alpha x} + (m - i e^{- \beta_{x}^{(0)} d z} \varphi_{x}^{(1)}) \bar{\psi}_{\alpha x} \psi_{\alpha x} \nn && + e^{- \alpha_{x}^{(0)} d z} ( \bar{\phi}_{\alpha x} p_{x}^{(0) \dagger} \psi_{\alpha x} + \bar{\psi}_{\alpha x} p_{x}^{(0)} \phi_{\alpha x}) - \int d^{D} x' \sqrt{g_{x'}^{(0)}} e^{- 2 \alpha_{x}^{(0)} d z} \Big[ \Big\{ \bar{\psi}_{\alpha x} \gamma^{c} e_{c x}^{\mu (0)} \Big( \partial_{\mu} - i q e^{- \delta_{x}^{(0)} d z} \tilde{a}_{\mu x}^{(1)} - \frac{i}{4} \omega_{\mu x}^{c'd' (0)} \sigma_{c'd'} \Big) \nn && + e^{\alpha_{x}^{(0)} d z} \bar{\phi}_{\alpha x} p_{x}^{(0) \dagger} \Big\} G_{xx'}^{(0)}[e_{a x}^{\nu (0)},\tilde{a}_{\nu x}^{(1)},\varphi_{x}^{(1)}] \Big] \Big[ e^{- 2 \alpha_{x'}^{(0)} d z} \gamma^{f} e_{f x'}^{\mu' (0)} \Big\{ \Big( \partial_{\mu'} - i q e^{- \delta_{x'}^{(0)} d z} \tilde{a}_{\mu' x'}^{(1)} - \frac{i}{4} \omega_{\mu' x'}^{f'g' (0)} \sigma_{f'g'} \Big) \psi_{\alpha x'} \nn && + e^{\alpha_{x'}^{(0)} d z} p_{x'}^{(0)} \phi_{\alpha x'} \Big\} \Big] + \frac{N}{2 \lambda_{\chi}} \varphi_{x}^{(0) 2} + \frac{N}{2 \lambda_{\chi}} \frac{1}{e^{2 \beta_{x}^{(0)} d z} - 1} (\varphi_{x}^{(1)} - \varphi_{x}^{(0)})^{2} \nn && + \frac{N q}{2 \lambda_{j}} g_{x}^{\mu\nu (0)} \Big(a_{\mu x}^{(0)} + \frac{i}{q} e^{\delta_{x}^{(0)} d z} d z \partial_{\mu} \alpha_{x}^{(0)} - e^{\delta_{x}^{(0)} d z} A_{B \mu x}\Big) \Big(a_{\nu x}^{(0)} + \frac{i}{q} e^{\delta_{x}^{(0)} d z} d z \partial_{\nu} \alpha_{x}^{(0)} - e^{\delta_{x}^{(0)} d z} A_{B \nu x}\Big) \nn && + \frac{N q}{2 \lambda_{j}} \frac{1}{e^{2 \delta_{x}^{(0)} d z} - 1} g_{x}^{\mu\nu (0)} (\tilde{a}_{\mu x}^{(1)} - a_{\mu x}^{(0)}) (\tilde{a}_{\nu x}^{(1)} - a_{\nu x}^{(0)}) - N \theta_{\mu x}^{a (0)} (e_{a x}^{\mu (0)} - e_{B a x}^{\mu}) - N [q_{x}^{(0) \dagger} (p_{x}^{(0)} - p_{B x}) + H.c.] \Bigg\} \Bigg] . \eqa
Although this expression is lengthy, it is easy to figure out how two terms are newly generated in this renormalization group transformation. The first newly generated term \bqa && N \mbox{tr}_{xx'} \ln \sqrt{g_{x}^{(0)}} e^{- 2 \alpha_{x}^{(0)} d z} \Big\{ \gamma^{a} e_{a x}^{\mu (0)} \Big( \partial_{\mu} - i q e^{- \delta_{x}^{(0)} d z} \tilde{a}_{\mu x}^{(1)} - \frac{i}{4} \omega_{\mu x}^{a'b' (0)} \sigma_{a'b'} \Big) + \frac{e^{2 \alpha_{x}^{(0)} d z} }{e^{2 \alpha_{x}^{(0)} d z} - 1} (m - i e^{- \beta_{x}^{(0)} d z} \varphi_{x}^{(1)}) \Big\} \nonumber \eqa results from $- \ln \int D \Psi_{\alpha} ~ e^{- \mathcal{S}_{\Psi\Psi}}$, where the effective action is \bqa \mathcal{S}_{\Psi\Psi} \equiv \int d^{D} x \sqrt{g^{(0)}} \Big\{ e^{- 2 \alpha^{(0)} d z} \bar{\Psi}_{\alpha} \gamma^{a} e_{a}^{\mu (0)} \Big( \partial_{\mu} - i q e^{- \delta^{(0)} d z} \tilde{a}_{\mu}^{(1)} - \frac{i}{4} \omega_{\mu}^{a'b' (0)} \sigma_{a'b'} \Big) \Psi_{\alpha} + \frac{(m - i e^{- \beta^{(0)} d z} \varphi^{(1)})}{e^{2 \alpha^{(0)} d z} - 1} \bar{\Psi}_{\alpha} \Psi_{\alpha} \Big\} . \nonumber \eqa This is nothing but the vacuum-fluctuation energy at zero temperature. The second newly generated term \bqa && \int d^{D} x \sqrt{g_{x}^{(0)}} \int d^{D} x' \sqrt{g_{x'}^{(0)}} e^{- 2 \alpha_{x}^{(0)} d z} \Big[ \Big\{ \bar{\psi}_{\alpha x} \gamma^{c} e_{c x}^{\mu (0)} \Big( \partial_{\mu} - i q e^{- \delta_{x}^{(0)} d z} \tilde{a}_{\mu x}^{(1)} - \frac{i}{4} \omega_{\mu x}^{c'd' (0)} \sigma_{c'd'} \Big) \nn && + e^{\alpha_{x}^{(0)} d z} \bar{\phi}_{\alpha x} p_{x}^{(0) \dagger} \Big\} G_{xx'}^{(0)}[e_{a x}^{\nu (0)},\tilde{a}_{\nu x}^{(1)},\varphi_{x}^{(1)}] \Big] \Big[ e^{- 2 \alpha_{x'}^{(0)} d z} \gamma^{f} e_{f x'}^{\mu' (0)} \Big\{ \Big( \partial_{\mu'} - i q e^{- \delta_{x'}^{(0)} d z} \tilde{a}_{\mu' x'}^{(1)} - \frac{i}{4} \omega_{\mu' x'}^{f'g' (0)} \sigma_{f'g'} \Big) \psi_{\alpha x'} \nn && + e^{\alpha_{x'}^{(0)} d z} p_{x'}^{(0)} \phi_{\alpha x'} \Big\} \Big] \nonumber \eqa
is given by \bqa && \Bigg\langle \exp\Bigg( - \int d^{D} x \sqrt{g^{(0)}} \Big[ e^{- 2 \alpha^{(0)} d z} \Big\{ \bar{\psi}_{\alpha} \gamma^{a} e_{a}^{\mu (0)} \Big( \partial_{\mu} - i q e^{- \delta^{(0)} d z} \tilde{a}_{\mu}^{(1)} - \frac{i}{4} \omega_{\mu}^{a'b' (0)} \sigma_{a'b'} \Big) + e^{\alpha^{(0)} d z} \bar{\phi}_{\alpha} p^{(0) \dagger} \Big\} \Psi_{\alpha} \nn && + e^{- 2 \alpha^{(0)} d z} \bar{\Psi}_{\alpha} \Big\{ \gamma^{a} e_{a}^{\mu (0)} \Big( \partial_{\mu} - i q e^{- \delta^{(0)} d z} \tilde{a}_{\mu}^{(1)} - \frac{i}{4} \omega_{\mu}^{a'b' (0)} \sigma_{a'b'} \Big) \psi_{\alpha} + e^{\alpha^{(0)} d z} p^{(0)} \phi_{\alpha} \Big\} \Big] \Bigg) \Bigg\rangle_{\Psi\Psi} , \nonumber \eqa
where $\Big\langle \mathcal{O}_{\Psi} \Big\rangle_{\Psi\Psi} \equiv \frac{1}{Z_{\Psi\Psi}} \int D \Psi_{\alpha} ~ \mathcal{O}_{\Psi} ~ e^{- \mathcal{S}_{\Psi\Psi}}$ with $Z_{\Psi\Psi} = \int D \Psi_{\alpha} ~ e^{- \mathcal{S}_{\Psi\Psi}}$. Here, $G_{xx'}^{(0)}[e_{a x}^{\nu (0)},\tilde{a}_{\nu x}^{(1)},\varphi_{x}^{(1)}]$ is the Green's function of the heavy Dirac-spinor field, given by
\bqa && \Big\{ e^{- 2 \alpha_{x}^{(0)} d z} \gamma^{a} e_{a x}^{\nu (0)} \Big( \partial_{\nu} - i q e^{- \delta_{x}^{(0)} d z} \tilde{a}_{\nu x}^{(1)} - \frac{i}{4} \omega_{\nu x}^{a'b' (0)} \sigma_{a'b'} \Big) + \frac{(m - i e^{- \beta_{x}^{(0)} d z} \varphi_{x}^{(1)})}{e^{2 \alpha_{x}^{(0)} d z} - 1} \Big\} G_{xx'}^{(0)}[e_{a x}^{\nu (0)},\tilde{a}_{\nu x}^{(1)},\varphi_{x}^{(1)}] \nn && = \frac{1}{\sqrt{g_{x}^{(0)}}} \delta^{(D)}(x-x') . \eqa
We point out that this correction is nonlocal in nature, clarified in the subscripts $x$ and $x'$. However, the heavy-mass nature given by $\sim (2 \alpha_{x}^{(0)} d z)^{-1}$ allows us to perform the gradient expansion, where the propagator is exponentially decaying. Below, we revisit this issue for renormalization of the kinetic-energy term.

To simplify further calculations, we choose the local speed of coarse graining as follows
\bqa && \partial_{\mu} \alpha_{x}^{(0)} = 0 , ~~~~~ \partial_{\mu} \beta_{x}^{(0)} = 0 , ~~~~~ \partial_{\mu} \delta_{x}^{(0)} = 0 . \eqa
Physically, this gauge choice implies that all the renormalization group transformations are performed with the same rate independent of spacetime. This gauge fixing leads us to lose general covariance. It turns out that this component plays the role of $g_{DD}$ in the emergent gravity description \cite{SungSik_Holography_I,SungSik_Holography_II,SungSik_Holography_III}, which corresponds to the metric tensor of this emergent extra-dimensional space.

Now, we perform the gradient expansion in $N \mbox{tr}_{xx'} \ln \sqrt{g_{x}^{(0)}} e^{- 2 \alpha^{(0)} d z} \Big\{ \gamma^{a} e_{a x}^{\mu (0)} \Big( \partial_{\mu} - i q e^{- \delta^{(0)} d z} \tilde{a}_{\mu x}^{(1)} - \frac{i}{4} \omega_{\mu x}^{a'b' (0)} \sigma_{a'b'} \Big) + \frac{e^{2 \alpha^{(0)} d z} }{e^{2 \alpha^{(0)} d z} - 1} (m - i e^{- \beta^{(0)} d z} \varphi_{x}^{(1)}) \Big\}$ for the background metric, U(1) gauge fields, and order parameter fields with respect to the uniform mass $m$ \cite{Gradient_Expansion_Gravity_I,Gradient_Expansion_Gravity_II}. Keeping all possible terms of the above partition function up to the linear order in $d z$, we obtain the following expression of the partition function
\bqa && Z[\phi_{\alpha x}, A_{B \mu x}] = \int D \psi_{\alpha x} D \varphi_{x}^{(0)} D \varphi_{x}^{(1)} D a_{\mu x}^{(0)} D \tilde{a}_{\mu x}^{(1)} D e_{a x}^{\mu (0)} D \theta^{a (0)}_{\mu x} D p_{x}^{(0)} D q_{x}^{(0)} \nn && \exp\Bigg[ - \int d^{D} x \sqrt{g_{x}^{(0)}} \Bigg\{ \bar{\psi}_{\alpha x} \gamma^{a} e_{a x}^{\mu (0)} \Big( \partial_{\mu} - i q \tilde{a}_{\mu x}^{(1)} - \frac{i}{4} \omega_{\mu x}^{a'b' (0)} \sigma_{a'b'} \Big) \psi_{\alpha x} + (m - i \varphi_{x}^{(1)}) \bar{\psi}_{\alpha x} \psi_{\alpha x} \nn && + ( \bar{\phi}_{\alpha x} p_{x}^{(0) \dagger} \psi_{\alpha x} + \bar{\psi}_{\alpha x} p_{x}^{(0)} \phi_{\alpha x}) - \int d^{D} x' \sqrt{g_{x'}^{(0)}} \Big[ \Big\{ \bar{\psi}_{\alpha x} \gamma^{c} e_{c x}^{\mu (0)} \Big( \partial_{\mu} - i q \tilde{a}_{\mu x}^{(1)} - \frac{i}{4} \omega_{\mu x}^{c'd' (0)} \sigma_{c'd'} \Big) \nn && + \bar{\phi}_{\alpha x} p_{x}^{(0) \dagger} \Big\} G_{xx'}^{(0)}[e_{a x}^{\nu (0)},\tilde{a}_{\nu x}^{(1)},\varphi_{x}^{(1)}] \Big] \Big[ \gamma^{f} e_{f x'}^{\mu' (0)} \Big\{ \Big( \partial_{\mu'} - i q \tilde{a}_{\mu' x'}^{(1)} - \frac{i}{4} \omega_{\mu' x'}^{f'g' (0)} \sigma_{f'g'} \Big) \psi_{\alpha x'} + p_{x'}^{(0)} \phi_{\alpha x'} \Big\} \Big] \nn && + \frac{N}{2 \lambda_{\chi}} \varphi_{x}^{(0) 2} + \frac{N q}{2 \lambda_{j}} g_{x}^{\mu\nu (0)} (a_{\mu x}^{(0)} - A_{B \mu x}) (a_{\nu x}^{(0)} - A_{B \nu x}) - N \theta_{\mu x}^{a (0)} (e_{a x}^{\mu (0)} - e_{B a x}^{\mu}) - N [q_{x}^{(0) \dagger} (p_{x}^{(0)} - p_{B x}) + H.c.] \nn && + 2 \beta^{(0)} d z \frac{N}{2 \lambda_{\chi}} \Big(\frac{\varphi_{x}^{(1)} - \varphi_{x}^{(0)}}{2 \beta^{(0)} d z}\Big)^{2} + 2 \delta^{(0)} d z \frac{N q}{2 \lambda_{j}} g_{x}^{\mu\nu (0)} \Big(\frac{\tilde{a}_{\mu x}^{(1)} - a_{\mu x}^{(0)}}{2 \delta^{(0)} d z}\Big) \Big(\frac{\tilde{a}_{\nu x}^{(1)} - a_{\nu x}^{(0)}}{2 \delta^{(0)} d z}\Big) \nn && + 2 \alpha^{(0)} d z N \Big( \frac{\mathcal{C}_{\varphi}}{2} g_{x}^{\mu\nu (0)} (\partial_{\mu} \varphi_{x}^{(1)}) (\partial_{\nu} \varphi_{x}^{(1)}) + \mathcal{C}_{\xi} R_{x}^{(0)} \varphi_{x}^{(1) 2} + \mathcal{C}_{mix} \tilde{F}_{x}^{\mu\nu (1)} (\partial_{\mu} \varphi_{x}^{(1)}) (\partial_{\nu} \varphi_{x}^{(1)}) + \mathcal{C}_{F} \tilde{F}_{\mu\nu x}^{(1)} \tilde{F}_{x}^{\mu\nu (1)} \nn && - \mathcal{C}_{\Lambda} + \mathcal{C}_{R} R_{x}^{(0)} \Big) \Bigg\} \Bigg] , \eqa
where the last line with $2 \alpha^{(0)} d z$ has been generated. Here, $\mathcal{C}_{\Lambda} = \frac{1}{2 \alpha^{(0)} d z} \ln \Big( \frac{m \sqrt{g_{x}^{(0)}}}{ 2 \alpha^{(0)} d z } \Big)$ denotes vacuum energy, identified with a cosmological constant. $R_{x}^{(0)}$ is Ricci scalar to give the kinetic energy for the metric tensor and $\mathcal{C}_{R}$ is a positive constant. $- \mathcal{C}_{\Lambda} + \mathcal{C}_{R} R_{x}$ is nothing but the Einstein-Hilbert action in $D$ spacetime dimensions, known to be the notion of induced gravity \cite{Gradient_Expansion_Gravity_I,Gradient_Expansion_Gravity_II}. $\tilde{F}_{\mu\nu x}^{(1)} \tilde{F}^{\mu\nu (1)}_{x}$ is the Maxwell action for the U(1) gauge field $\tilde{a}_{\mu x}^{(1)}$ and $\mathcal{C}_{F}$ is a positive constant. The kinetic-energy term for order-parameter fluctuations $\varphi_{x}^{(1)}$ is generated from vacuum fluctuations, where $\mathcal{C}_{\varphi}$ is the stiffness parameter. There also appear coupling terms between the scalar curvature and the density of order-parameter fields and between the field strength tensor of U(1) gauge fields and the kinetic energy tensor of order-parameter fields with their positive coefficients $\mathcal{C}_{\xi}$ and $\mathcal{C}_{mix}$, respectively. All the coefficients of $\mathcal{C}_{R}$, $\mathcal{C}_{F}$, $\mathcal{C}_{\varphi}$, $\mathcal{C}_{\xi}$, and $\mathcal{C}_{mix}$ decrease as the bare mass $m$ of Dirac spinor fields increases. There may exist possible anomaly terms involved with gravitational curvature, electromagnetic field strength, and their mixing, respectively. Here, we do not consider such anomaly terms.
%
%
The Green's function of the heavy Dirac spinor is given by
\bqa && \Big\{ \gamma^{a} e_{a x}^{\nu (0)} \Big( \partial_{\nu} - i q \tilde{a}_{\nu x}^{(1)} - \frac{i}{4} \omega_{\nu x}^{a'b' (0)} \sigma_{a'b'} \Big) + \frac{1}{ 2 \alpha^{(0)} d z } (m - i \varphi_{x}^{(1)}) \Big\} G_{xx'}^{(0)}[e_{a x}^{\nu (0)},\tilde{a}_{\nu x}^{(1)},\varphi_{x}^{(1)}] = \frac{1}{\sqrt{g_{x}^{(0)}}} \delta^{(D)}(x-x') . \nn \eqa
This partition function is the result of the first renormalization group transformation.

%
%
%

\subsection{Recursive renormalization group transformations}

\subsubsection{Locality approximation for the Green's function}

To implement the second renormalization group transformation with respect to $\varphi^{(1)}$, $\tilde{a}_{\mu}^{(1)}$, and $\psi_{\alpha}$, it is necessary to rewrite the effective action of the partition function in the same form as the original expression of the Dirac Lagrangian, where both the vierbein and U(1) gauge fields have to be renormalized and redefined appropriately. First of all, we keep the leading order in the gradient expansion for the Green's function, where higher order derivatives are given by higher powers of $d z$, thus safely neglected in the $d z \rightarrow 0$ limit. We call this approximation the locality approximation of the Green's function. Within the locality approximation, the partition function is given by
\bqa && Z[\phi_{\alpha x}, A_{B \mu x}] = \int D \psi_{\alpha x} D \varphi_{x}^{(0)} D \varphi_{x}^{(1)} D a_{\mu x}^{(0)} D \tilde{a}_{\mu x}^{(1)} D e_{a x}^{\mu (0)} D \theta^{a (0)}_{\mu x} D p_{x}^{(0)} D q_{x}^{(0)} \nn && \exp\Bigg[ - \int d^{D} x \sqrt{g_{x}^{(0)}} \Bigg\{ \bar{\psi}_{\alpha x} \Big[ 1 - \Big\{ \gamma^{c} e_{c x}^{\nu (0)} \Big( \partial_{\nu} - i q \tilde{a}_{\nu x}^{(1)} - \frac{i}{4} \omega_{\nu x}^{c'd' (0)} \sigma_{c'd'} \Big) G_{xx'}^{(0)}[e_{a x}^{\nu (0)},\tilde{a}_{\nu x}^{(1)},\varphi_{x}^{(1)}] \Big\}_{x' \rightarrow x} \Big] \gamma^{a} e_{a x}^{\mu (0)} \Big( \partial_{\mu} \nn && - i q \tilde{a}_{\mu x}^{(1)} - \frac{i}{4} \omega_{\mu x}^{a'b' (0)} \sigma_{a'b'} \Big) \psi_{\alpha x} + (m - i \varphi_{x}^{(1)}) \bar{\psi}_{\alpha x} \psi_{\alpha x} + \bar{\phi}_{\alpha x} p_{x}^{(0) \dagger} \Big\{ 1 - \Big( G_{xx'}^{(0)}[e_{a x}^{\nu (0)},\tilde{a}_{\nu x}^{(1)},\varphi_{x}^{(1)}] \Big)_{x' \rightarrow x} \gamma^{a} e_{a x}^{\mu (0)} \Big( \partial_{\mu} \nn && - i q \tilde{a}_{\mu x}^{(1)} - \frac{i}{4} \omega_{\mu x}^{a'b' (0)} \sigma_{a'b'} \Big) \Big\} \psi_{\alpha x} + \bar{\psi}_{\alpha x} \Big[ 1 - \Big\{ \gamma^{c} e_{c x}^{\mu (0)} \Big( \partial_{\mu} - i q \tilde{a}_{\mu x}^{(1)} - \frac{i}{4} \omega_{\mu x}^{c'd' (0)} \sigma_{c'd'} \Big) G_{xx'}^{(0)}[e_{a x}^{\nu (0)},\tilde{a}_{\nu x}^{(1)},\varphi_{x}^{(1)}] \Big\}_{x' \rightarrow x} \Big] p_{x}^{(0)} \phi_{\alpha x} \nn && - \bar{\phi}_{\alpha x} p_{x}^{(0) \dagger} \Big( G_{xx'}^{(0)}[e_{a x}^{\nu (0)},\tilde{a}_{\nu x}^{(1)},\varphi_{x}^{(1)}] \Big)_{x' \rightarrow x} p_{x}^{(0)} \phi_{\alpha x} \nn && + \frac{N}{2 \lambda_{\chi}} \varphi_{x}^{(0) 2} + \frac{N q}{2 \lambda_{j}} g_{x}^{\mu\nu (0)} (a_{\mu x}^{(0)} - A_{B \mu x}) (a_{\nu x}^{(0)} - A_{B \nu x}) - N \theta_{\mu x}^{a (0)} (e_{a x}^{\mu (0)} - e_{B a x}^{\mu}) - N [q_{x}^{(0) \dagger} (p_{x}^{(0)} - p_{B x}) + H.c.] \nn && + 2 \beta^{(0)} d z \frac{N}{2 \lambda_{\chi}} \Big(\frac{\varphi_{x}^{(1)} - \varphi_{x}^{(0)}}{2 \beta^{(0)} d z}\Big)^{2} + 2 \delta^{(0)} d z \frac{N q}{2 \lambda_{j}} g_{x}^{\mu\nu (0)} \Big(\frac{\tilde{a}_{\mu x}^{(1)} - a_{\mu x}^{(0)}}{2 \delta^{(0)} d z}\Big) \Big(\frac{\tilde{a}_{\nu x}^{(1)} - a_{\nu x}^{(0)}}{2 \delta^{(0)} d z}\Big) \nn && + 2 \alpha^{(0)} d z N \Big( \frac{\mathcal{C}_{\varphi}}{2} g_{x}^{\mu\nu (0)} (\partial_{\mu} \varphi_{x}^{(1)}) (\partial_{\nu} \varphi_{x}^{(1)}) + \mathcal{C}_{\xi} R_{x}^{(0)} \varphi_{x}^{(1) 2} + \mathcal{C}_{mix} \tilde{F}_{x}^{\mu\nu (1)} (\partial_{\mu} \varphi_{x}^{(1)}) (\partial_{\nu} \varphi_{x}^{(1)}) + \mathcal{C}_{F} \tilde{F}_{\mu\nu x}^{(1)} \tilde{F}_{x}^{\mu\nu (1)} \nn && - \mathcal{C}_{\Lambda} + \mathcal{C}_{R} R_{x}^{(0)} \Big) \Bigg\} \Bigg] . \eqa
We emphasize that the renormalization group transformation given by quantum fluctuations of high-energy Dirac fermions results in the renormalization of the kinetic-energy term for low-energy Dirac fermions, expressed as $\Big[ 1 - \Big\{ \gamma^{c} e_{c x}^{\nu (0)} \Big( \partial_{\nu} - i q \tilde{a}_{\nu x}^{(1)} - \frac{i}{4} \omega_{\nu x}^{c'd' (0)} \sigma_{c'd'} \Big) G_{xx'}^{(0)}[e_{a x}^{\nu (0)},\tilde{a}_{\nu x}^{(1)},\varphi_{x}^{(1)}] \Big\}_{x' \rightarrow x} \Big]$ in the kinetic-energy term. We point out that this renormalization structure is universal regardless of quantum field theories of bosons or fermions, where the high-energy Green's function of original matter fields encodes the renormalization group transformation of the kinetic-energy term of the corresponding quantum field theory \cite{Holographic_Description_Kim_Ryu}. This renormalization group transformation also gives rise to the renormalization of the source term. In addition, $- \bar{\phi}_{\alpha x} p_{x}^{(0) \dagger} \Big( G_{xx'}^{(0)}[e_{a x}^{\nu (0)},\tilde{a}_{\nu x}^{(1)},\varphi_{x}^{(1)}] \Big)_{x' \rightarrow x} p_{x}^{(0)} \phi_{\alpha x}$ is newly generated in the effective Lagrangian.

%
%
%

\subsubsection{Update rules}

A key observation is that we can exponentiate all the terms including the Green's function of heavy Dirac fermions since it is of the order of $d z$. As a result, we reformulate the previous partition function up to the linear order of $d z$ in the following way
\bqa && Z[\phi_{\alpha x}, A_{B \mu x}] = \int D \psi_{\alpha x} D \varphi_{x}^{(0)} D \varphi_{x}^{(1)} D a_{\mu x}^{(0)} D \tilde{a}_{\mu x}^{(1)} D e_{a x}^{\mu (0)} D \theta^{a (0)}_{\mu x} D p_{x}^{(0)} D q_{x}^{(0)} \nn && \exp\Bigg[ - \int d^{D} x \sqrt{g_{x}^{(0)}} \Bigg\{ \bar{\psi}_{\alpha x} e^{- \Big\{ \gamma^{c} e_{c x}^{\nu (0)} \Big( \partial_{\nu} - i q \tilde{a}_{\nu x}^{(1)} - \frac{i}{4} \omega_{\nu x}^{c'd' (0)} \sigma_{c'd'} \Big) G_{xx'}^{(0)}[e_{a x}^{\nu (0)},\tilde{a}_{\nu x}^{(1)},\varphi_{x}^{(1)}] \Big\}_{x' \rightarrow x}} \gamma^{a} e_{a x}^{\mu (0)} \Big( \partial_{\mu} - i q \tilde{a}_{\mu x}^{(1)} \nn && - \frac{i}{4} \omega_{\mu x}^{a'b' (0)} \sigma_{a'b'} \Big) \psi_{\alpha x} + (m - i \varphi_{x}^{(1)}) \bar{\psi}_{\alpha x} \psi_{\alpha x} + \bar{\phi}_{\alpha x} p_{x}^{(0) \dagger} e^{- \Big( G_{xx'}^{(0)}[e_{a x}^{\nu (0)},\tilde{a}_{\nu x}^{(1)},\varphi_{x}^{(1)}] \Big)_{x' \rightarrow x} \gamma^{a} e_{a x}^{\mu (0)} \Big( \partial_{\mu} - i q \tilde{a}_{\mu x}^{(1)} - \frac{i}{4} \omega_{\mu x}^{a'b' (0)} \sigma_{a'b'} \Big)} \psi_{\alpha x} \nn && + \bar{\psi}_{\alpha x} e^{- \Big\{ \gamma^{c} e_{c x}^{\mu (0)} \Big( \partial_{\mu} - i q \tilde{a}_{\mu x}^{(1)} - \frac{i}{4} \omega_{\mu x}^{c'd' (0)} \sigma_{c'd'} \Big) G_{xx'}^{(0)}[e_{a x}^{\nu (0)},\tilde{a}_{\nu x}^{(1)},\varphi_{x}^{(1)}] \Big\}_{x' \rightarrow x}} p_{x}^{(0)} \phi_{\alpha x} - \bar{\phi}_{\alpha x} p_{x}^{(0) \dagger} \Big( G_{xx'}^{(0)}[e_{a x}^{\nu (0)},\tilde{a}_{\nu x}^{(1)},\varphi_{x}^{(1)}] \Big)_{x' \rightarrow x} p_{x}^{(0)} \phi_{\alpha x} \nn && + \frac{N}{2 \lambda_{\chi}} \varphi_{x}^{(0) 2} + \frac{N q}{2 \lambda_{j}} g_{x}^{\mu\nu (0)} (a_{\mu x}^{(0)} - A_{B \mu x}) (a_{\nu x}^{(0)} - A_{B \nu x}) - N \theta_{\mu x}^{a (0)} (e_{a x}^{\mu (0)} - e_{B a x}^{\mu}) - N [q_{x}^{(0) \dagger} (p_{x}^{(0)} - p_{B x}) + H.c.] \nn && + 2 \beta^{(0)} d z \frac{N}{2 \lambda_{\chi}} \Big(\frac{\varphi_{x}^{(1)} - \varphi_{x}^{(0)}}{2 \beta^{(0)} d z}\Big)^{2} + 2 \delta^{(0)} d z \frac{N q}{2 \lambda_{j}} g_{x}^{\mu\nu (0)} \Big(\frac{\tilde{a}_{\mu x}^{(1)} - a_{\mu x}^{(0)}}{2 \delta^{(0)} d z}\Big) \Big(\frac{\tilde{a}_{\nu x}^{(1)} - a_{\nu x}^{(0)}}{2 \delta^{(0)} d z}\Big) \nn && + 2 \alpha^{(0)} d z N \Big( \frac{\mathcal{C}_{\varphi}}{2} g_{x}^{\mu\nu (0)} (\partial_{\mu} \varphi_{x}^{(1)}) (\partial_{\nu} \varphi_{x}^{(1)}) + \mathcal{C}_{\xi} R_{x}^{(0)} \varphi_{x}^{(1) 2} + \mathcal{C}_{mix} \tilde{F}_{x}^{\mu\nu (1)} (\partial_{\mu} \varphi_{x}^{(1)}) (\partial_{\nu} \varphi_{x}^{(1)}) + \mathcal{C}_{F} \tilde{F}_{\mu\nu x}^{(1)} \tilde{F}_{x}^{\mu\nu (1)} \nn && - \mathcal{C}_{\Lambda} + \mathcal{C}_{R} R_{x}^{(0)} \Big) \Bigg\} \Bigg] . \eqa

Now, it is straightforward to find the update rule of the vierbein tensor. Our object is to reformulate the kinetic-energy term as follows
\bqa && e^{- \Big\{ \gamma^{c} e_{c x}^{\nu (0)} \Big( \partial_{\nu} - i q \tilde{a}_{\nu x}^{(1)} - \frac{i}{4} \omega_{\nu x}^{c'd' (0)} \sigma_{c'd'} \Big) G_{xx'}^{(0)}[e_{a x}^{\nu (0)},\tilde{a}_{\nu x}^{(1)},\varphi_{x}^{(1)}] \Big\}_{x' \rightarrow x}} \gamma^{a} e_{a x}^{\mu (0)} \Big( \partial_{\mu} - i q \tilde{a}_{\mu x}^{(1)} - \frac{i}{4} \omega_{\mu x}^{a'b' (0)} \sigma_{a'b'} \Big) \nn && \equiv \gamma^{a} e_{a x}^{\mu (1)} \Big( \partial_{\mu} - i q a_{\mu x}^{(1)} - \frac{i}{4} \omega_{\mu x}^{a'b' (1)} \sigma_{a'b'} \Big) . \eqa
Keeping all terms up to the leading order of $d z$, the renormalized vierbein tensor is given by
\bqa && e_{a x}^{\mu (1)} = e_{a x}^{\mu (0)} - \frac{1}{D} \mbox{tr} \Big[ \gamma_{a} \Big\{ \gamma^{c} e_{c x}^{\nu (0)} \Big( \partial_{\nu} - i q \tilde{a}_{\nu x}^{(1)} - \frac{i}{4} \omega_{\nu x}^{c'd' (0)} \sigma_{c'd'} \Big) G_{xx'}^{(0)}[e_{a x}^{\nu (0)},\tilde{a}_{\nu x}^{(1)},\varphi_{x}^{(1)}] \Big\}_{x' \rightarrow x} \gamma^{b} \Big] e_{b x}^{\mu (0)} . \label{Vierbein_Update} \eqa
Accordingly, the renormalized spin connection follows
\bqa && \omega_{\mu x}^{a'b' (1)} \sigma_{a'b'} = \omega_{\mu x}^{a'b' (0)} \sigma_{a'b'} \nn && - \frac{1}{D} e^{a (0)}_{\mu x} \mbox{tr} \Big[ \gamma_{a} \Big\{ \gamma^{c} e_{c x}^{\nu (0)} \Big( \partial_{\nu} - i q \tilde{a}_{\nu x}^{(1)} - \frac{i}{4} \omega_{\nu x}^{c'd' (0)} \sigma_{c'd'} \Big) G_{xx'}^{(0)}[e_{a x}^{\nu (0)},\tilde{a}_{\nu x}^{(1)},\varphi_{x}^{(1)}] \Big\}_{x' \rightarrow x} \gamma^{f} \Big] e_{f x}^{\lambda (0)} \omega_{\lambda x}^{f'g' (0)} \sigma_{f'g'} . \eqa
In the same way, the U(1) gauge field is updated as
\bqa && a_{\mu x}^{(1)} = \tilde{a}_{\mu x}^{(1)} - \frac{1}{D} e^{a (0)}_{\mu x} \mbox{tr} \Big[ \gamma_{a} \Big\{ \gamma^{c} e_{c x}^{\nu (0)} \Big( \partial_{\nu} - i q \tilde{a}_{\nu x}^{(1)} - \frac{i}{4} \omega_{\nu x}^{c'd' (0)} \sigma_{c'd'} \Big) G_{xx'}^{(0)}[e_{a x}^{\nu (0)},\tilde{a}_{\nu x}^{(1)},\varphi_{x}^{(1)}] \Big\}_{x' \rightarrow x} \gamma^{b} \Big] e_{b x}^{\lambda (0)} \tilde{a}_{\lambda x}^{(1)} . \label{Gauge_Field_Update} \eqa

The coupling constant of the source-field term is renormalized as
\bqa && p_{x}^{(1)} = e^{- \Big\{ \gamma^{c} e_{c x}^{\mu (0)} \Big( \partial_{\mu} - i q \tilde{a}_{\mu x}^{(1)} - \frac{i}{4} \omega_{\mu x}^{c'd' (0)} \sigma_{c'd'} \Big) G_{xx'}^{(0)}[e_{a x}^{\nu (0)},\tilde{a}_{\nu x}^{(1)},\varphi_{x}^{(1)}] \Big\}_{x' \rightarrow x}} p_{x}^{(0)} , \eqa
which can be translated into
\bqa && p_{x}^{(1)} = p_{x}^{(0)} - \Big\{ \gamma^{c} e_{c x}^{\mu (0)} \Big( \partial_{\mu} - i q \tilde{a}_{\mu x}^{(1)} - \frac{i}{4} \omega_{\mu x}^{c'd' (0)} \sigma_{c'd'} \Big) G_{xx'}^{(0)}[e_{a x}^{\nu (0)},\tilde{a}_{\nu x}^{(1)},\varphi_{x}^{(1)}] \Big\}_{x' \rightarrow x} p_{x}^{(0)} \label{Source_Term_Update} \eqa
up to the linear order of $d z$.

Based on the above update procedure, we reach the following expression of the partition function after the first renormalization group transformation
\bqa && Z[\phi_{\alpha x}, A_{B \mu x}] \nn && = \int D \psi_{\alpha x} D \varphi_{x}^{(0)} D \varphi_{x}^{(1)} D a_{\mu x}^{(0)} D a_{\mu x}^{(1)} D e_{a x}^{\mu (0)} D e_{a x}^{\mu (1)} D \theta^{a (0)}_{\mu x} D \theta^{a (1)}_{\mu x} D p_{x}^{(0)} D p_{x}^{(1)} D q_{x}^{(0)} D q_{x}^{(1)} D y_{x}^{(0)} D y_{x}^{(1)} D w_{x}^{(0)} D w_{x}^{(1)} \nn && \exp\Bigg[ - \int d^{D} x \sqrt{g_{x}^{(1)}} \Bigg\{ \bar{\psi}_{\alpha x} \gamma^{a} e_{a x}^{\mu (1)} \Big( \partial_{\mu} - i q a_{\mu x}^{(1)} - \frac{i}{4} \omega_{\mu x}^{a'b' (1)} \sigma_{a'b'} \Big) \psi_{\alpha x} + (m - i \varphi_{x}^{(1)}) \bar{\psi}_{\alpha x} \psi_{\alpha x} \nn && + \bar{\phi}_{\alpha x} p_{x}^{(1) \dagger} \psi_{\alpha x} + \bar{\psi}_{\alpha x} p_{x}^{(1)} \phi_{\alpha x} + \bar{\phi}_{\alpha x} y_{x}^{(1)} \phi_{\alpha x} \Bigg\} - N \int d^{D} x \sqrt{g_{x}^{(0)}} \Bigg\{ \frac{1}{2 \lambda_{\chi}} \varphi_{x}^{(0) 2} + \frac{q}{2 \lambda_{j}} g_{x}^{\mu\nu (0)} (a_{\mu x}^{(0)} - A_{B \mu x}) (a_{\nu x}^{(0)} - A_{B \nu x}) \nn && - \theta_{\mu x}^{a (0)} (e_{a x}^{\mu (0)} - e_{B a x}^{\mu}) - q_{x}^{(0) \dagger} (p_{x}^{(0)} - p_{B x}) - H.c. - w_{x}^{(0) \dagger} (y_{x}^{(0)} - y_{B x}) - H.c. \Bigg\} \nn && - 2 d z N \int d^{D} x \sqrt{g_{x}^{(0)}} \Bigg\{ \frac{1}{2 \lambda_{\chi}} \Big(\frac{\varphi_{x}^{(1)} - \varphi_{x}^{(0)}}{2 d z}\Big)^{2} + \frac{q}{2 \lambda_{j}} g_{x}^{\mu\nu (0)} \Bigg(\frac{a_{\mu x}^{(1)} - a_{\mu x}^{(0)}}{2 d z} \nn && + \frac{1}{2 d z D} e^{a (0)}_{\mu x} \mbox{tr} \Big[ \gamma_{a} \Big\{ \gamma^{c} e_{c x}^{\lambda (0)} \Big( \partial_{\lambda} - i q a_{\lambda x}^{(1)} - \frac{i}{4} \omega_{\lambda x}^{c'd' (0)} \sigma_{c'd'} \Big) G_{xx'}^{(0)}[e_{a x}^{\lambda (0)},a_{\lambda x}^{(1)},\varphi_{x}^{(1)}] \Big\}_{x' \rightarrow x} \gamma^{b} \Big] e_{b x}^{\kappa (0)} a_{\kappa x}^{(1)}\Bigg) \Bigg(\frac{a_{\nu x}^{(1)} - a_{\nu x}^{(0)}}{2 d z} \nn && + \frac{1}{2 d z D} e^{f (0)}_{\nu x} \mbox{tr} \Big[ \gamma_{f} \Big\{ \gamma^{g} e_{g x}^{\eta (0)} \Big( \partial_{\eta} - i q a_{\eta x}^{(1)} - \frac{i}{4} \omega_{\eta x}^{g'h' (0)} \sigma_{g'h'} \Big) G_{xx'}^{(0)}[e_{a x}^{\eta (0)},a_{\eta x}^{(1)},\varphi_{x}^{(1)}] \Big\}_{x' \rightarrow x} \gamma^{l} \Big] e_{l x}^{\epsilon (0)} a_{\epsilon x}^{(1)}\Bigg) \nn && + \frac{\mathcal{C}_{\varphi}}{2} g_{x}^{\mu\nu (0)} (\partial_{\mu} \varphi_{x}^{(1)}) (\partial_{\nu} \varphi_{x}^{(1)}) + \mathcal{C}_{\xi} R_{x}^{(0)} \varphi_{x}^{(1) 2} + \mathcal{C}_{mix} F_{x}^{\mu\nu (1)} (\partial_{\mu} \varphi_{x}^{(1)}) (\partial_{\nu} \varphi_{x}^{(1)}) + \mathcal{C}_{F} F_{\mu\nu x}^{(1)} F_{x}^{\mu\nu (1)} - \mathcal{C}_{\Lambda} + \mathcal{C}_{R} R_{x}^{(0)} \Bigg\} \nn && - 2 d z N \int d^{D} x \sqrt{g_{x}^{(1)}} \Bigg\{ - \theta_{\mu x}^{a (1)} \Bigg( \frac{e_{a x}^{\mu (1)} - e_{a x}^{\mu (0)}}{2 d z} \nn && + \frac{1}{2 d z D} \mbox{tr} \Big[ \gamma_{a} \Big\{ \gamma^{c} e_{c x}^{\nu (0)} \Big( \partial_{\nu} - i q a_{\nu x}^{(1)} - \frac{i}{4} \omega_{\nu x}^{c'd' (0)} \sigma_{c'd'} \Big) G_{xx'}^{(0)}[e_{a x}^{\nu (0)},a_{\nu x}^{(1)},\varphi_{x}^{(1)}] \Big\}_{x' \rightarrow x} \gamma^{b} \Big] e_{b x}^{\mu (0)} \Bigg) \nn && - q_{x}^{(1) \dagger} \Bigg( \frac{p_{x}^{(1)} - p_{x}^{(0)}}{2 d z} + \frac{1}{2 d z} \Big\{ \gamma^{c} e_{c x}^{\mu (0)} \Big( \partial_{\mu} - i q a_{\mu x}^{(1)} - \frac{i}{4} \omega_{\mu x}^{c'd' (0)} \sigma_{c'd'} \Big) G_{xx'}^{(0)}[e_{a x}^{\nu (0)},a_{\nu x}^{(1)},\varphi_{x}^{(1)}] \Big\}_{x' \rightarrow x} p_{x}^{(0)} \Bigg) - H.c. \nn && - w_{x}^{(1) \dagger} \Bigg( \frac{y_{x}^{(1)} - y_{x}^{(0)}}{2 d z} + \frac{1}{2 d z} p_{x}^{(0) \dagger} \Big( G_{xx'}^{(0)}[e_{a x}^{\nu (0)},a_{\nu x}^{(1)},\varphi_{x}^{(1)}] \Big)_{x' \rightarrow x} p_{x}^{(0)} \Bigg) - H.c. \Bigg\} \Bigg] , \eqa
where the gauge fixing of \bqa && \alpha^{(0)} = \beta^{(0)} = \delta^{(0)} = 1 \eqa has been utilized. The Green's function is
\bqa && \Big\{ \gamma^{a} e_{a x}^{\nu (0)} \Big( \partial_{\nu} - i q a_{\nu x}^{(1)} - \frac{i}{4} \omega_{\nu x}^{a'b' (0)} \sigma_{a'b'} \Big) + \frac{1}{ 2 d z } (m - i \varphi_{x}^{(1)}) \Big\} G_{xx'}^{(0)}[e_{a x}^{\nu (0)},a_{\nu x}^{(1)},\varphi_{x}^{(1)}] = \frac{1}{\sqrt{g_{x}^{(0)}}} \delta^{(D)}(x-x') , \eqa
where the U(1) gauge field has been updated and verified in the $d z \rightarrow 0$ limit.

This lengthy expression is not difficult to figure out. First of all, both the vierbein tensor and the U(1) gauge field are renormalized in the Dirac Lagrangian. The renormalization group flow of the vierbein tensor is described by \bqa && - \theta_{\mu x}^{a (1)} \Bigg( \frac{e_{a x}^{\mu (1)} - e_{a x}^{\mu (0)}}{2 d z} + \frac{1}{2 d z D} \mbox{tr} \Big[ \gamma_{a} \Big\{ \gamma^{c} e_{c x}^{\nu (0)} \Big( \partial_{\nu} - i q a_{\nu x}^{(1)} - \frac{i}{4} \omega_{\nu x}^{c'd' (0)} \sigma_{c'd'} \Big) G_{xx'}^{(0)}[e_{a x}^{\nu (0)},a_{\nu x}^{(1)},\varphi_{x}^{(1)}] \Big\}_{x' \rightarrow x} \gamma^{b} \Big] e_{b x}^{\mu (0)} \Bigg) , \nonumber \eqa where $\theta_{\mu x}^{a (1)}$ is the Lagrange multiplier field to impose the constraint of Eq. (\ref{Vierbein_Update}). We recall that the Einstein-Hilbert action $- \mathcal{C}_{\Lambda} + \mathcal{C}_{R} R_{x}^{(0)}$ originates from vacuum fluctuations of high-energy Dirac fermions. On the other hand, the dynamics of renormalized U(1) gauge fields is given by \bqa && \mathcal{L}_{g}^{(1)} = \frac{q g_{x}^{\mu\nu (0)}}{2 \lambda_{j}} \Bigg(\frac{a_{\mu x}^{(1)} - a_{\mu x}^{(0)}}{2 d z} + \frac{1}{2 d z D} e^{a (0)}_{\mu x} \mbox{tr} \Big[ \gamma_{a} \Big\{ \gamma^{c} e_{c x}^{\lambda (0)} \Big( \partial_{\lambda} - i q a_{\lambda x}^{(1)} - \frac{i}{4} \omega_{\lambda x}^{c'd' (0)} \sigma_{c'd'} \Big) G_{xx'}^{(0)}[e_{a x}^{\lambda (0)},a_{\lambda x}^{(1)},\varphi_{x}^{(1)}] \Big\}_{x' \rightarrow x} \gamma^{b} \Big] \nn && \times e_{b x}^{\kappa (0)} a_{\kappa x}^{(1)}\Bigg) \Bigg(\frac{a_{\nu x}^{(1)} - a_{\nu x}^{(0)}}{2 d z} + \frac{1}{2 d z D} e^{f (0)}_{\nu x} \mbox{tr} \Big[ \gamma_{f} \Big\{ \gamma^{g} e_{g x}^{\eta (0)} \Big( \partial_{\eta} - i q a_{\eta x}^{(1)} - \frac{i}{4} \omega_{\eta x}^{g'h' (0)} \sigma_{g'h'} \Big) G_{xx'}^{(0)}[e_{a x}^{\eta (0)},a_{\eta x}^{(1)},\varphi_{x}^{(1)}] \Big\}_{x' \rightarrow x} \gamma^{l} \Big] \nn && \times e_{l x}^{\epsilon (0)} a_{\epsilon x}^{(1)}\Bigg) + \mathcal{C}_{F} F_{\mu\nu x}^{(1)} F_{x}^{\mu\nu (1)} , \nonumber \eqa
where the dynamics $g_{x}^{\mu\nu (0)} \Big(\frac{a_{\mu x}^{(1)} - a_{\mu x}^{(0)}}{2 d z} \Big) \Big(\frac{a_{\nu x}^{(1)} - a_{\nu x}^{(0)}}{2 d z}\Big)$ originates from effective current-current interactions and the Maxwell dynamics $F_{\mu\nu x}^{(1)} F_{x}^{\mu\nu (1)}$ results from quantum fluctuations of high-energy Dirac fermions. The mass term of Dirac fermions is also renormalized as $(m - i \varphi_{x}^{(1)}) \bar{\psi}_{\alpha x} \psi_{\alpha x}$, where the dynamics of the order-parameter field is given by \bqa && \mathcal{L}_{\varphi}^{(1)} = \frac{1}{2 \lambda_{\chi}} \Big(\frac{\varphi_{x}^{(1)} - \varphi_{x}^{(0)}}{2 d z}\Big)^{2} + \frac{\mathcal{C}_{\varphi}}{2} g_{x}^{\mu\nu (0)} (\partial_{\mu} \varphi_{x}^{(1)}) (\partial_{\nu} \varphi_{x}^{(1)}) + \mathcal{C}_{\xi} R_{x}^{(0)} \varphi_{x}^{(1) 2} + \mathcal{C}_{mix} F_{x}^{\mu\nu (1)} (\partial_{\mu} \varphi_{x}^{(1)}) (\partial_{\nu} \varphi_{x}^{(1)}) . \nonumber \eqa
The dynamics of $\frac{1}{2 \lambda_{\chi}} \Big(\frac{\varphi_{x}^{(1)} - \varphi_{x}^{(0)}}{2 d z}\Big)^{2}$ originates from effective interactions of Dirac fermions for dynamical mass generation and all other terms result from vacuum fluctuations of high-energy Dirac fermions. We point out that the coupling function of the source term introduced for the Green's function is renormalized, given by \bqa && - q_{x}^{(1) \dagger} \Bigg( \frac{p_{x}^{(1)} - p_{x}^{(0)}}{2 d z} + \frac{1}{2 d z} \Big\{ \gamma^{c} e_{c x}^{\mu (0)} \Big( \partial_{\mu} - i q a_{\mu x}^{(1)} - \frac{i}{4} \omega_{\mu x}^{c'd' (0)} \sigma_{c'd'} \Big) G_{xx'}^{(0)}[e_{a x}^{\nu (0)},a_{\nu x}^{(1)},\varphi_{x}^{(1)}] \Big\}_{x' \rightarrow x} p_{x}^{(0)} \Bigg) - H.c. , \nonumber \eqa where $q_{x}^{(1) \dagger}$ is the Lagrange multiplier field to impose the constraint Eq. (\ref{Source_Term_Update}). The renormalization group flow of the newly generated source term is described by \bqa && - w_{x}^{(1) \dagger} \Bigg( \frac{y_{x}^{(1)} - y_{x}^{(0)}}{2 d z} + \frac{1}{2 d z} p_{x}^{(0) \dagger} \Big( G_{xx'}^{(0)}[e_{a x}^{\nu (0)},a_{\nu x}^{(1)},\varphi_{x}^{(1)}] \Big)_{x' \rightarrow x} p_{x}^{(0)} \Bigg) - H.c. , \nonumber \eqa where $w_{x}^{(1) \dagger}$ is the Lagrange multiplier field and $y_{x}^{(0)} = 0$ as an initial condition.

Physical ingredients in the above partition function are as follows. It is easy to figure out that quantum fluctuations of $\varphi_{x}^{(0)}$, $\varphi_{x}^{(1)}$, $a_{\mu x}^{(0)}$, and $a_{\mu x}^{(1)}$ are suppressed in the large $N$ limit while the path integral for $\psi_{\alpha x}$ is Gaussian and those for both vierbein fields are given by the constraints of $\delta-$functions. As a result, we obtain effective mean-field equations for $\varphi_{x}^{(0)}$, $\varphi_{x}^{(1)}$, $a_{\mu x}^{(0)}$, and $a_{\mu x}^{(1)}$ with background geometry given by the vierbein tensor in the large $N$ limit. Mean-field equations of $\varphi_{x}^{(0)}$ and $a_{\mu x}^{(0)}$ correspond to those (conventional saddle-point approximation) of the related quantum field theory in the large $N$ limit. On the other hand, mean-field equations of $\varphi_{x}^{(1)}$ and $a_{\mu x}^{(1)}$ coupled to those of $\varphi_{x}^{(0)}$ and $a_{\mu x}^{(0)}$ give rise to $1/N$ corrections in the mean-field solutions of $\varphi_{x}^{(0)}$ and $a_{\mu x}^{(0)}$. This point has been demonstrated in Ref. \cite{Holographic_Liquid_Kim} more explicitly. In addition, the renormalized vierbein tensor corresponds to vertex corrections, where all coupling functions are renormalized. In this respect the above effective action of the large $N$ limit is an effective mean-field theory for $\varphi_{x}^{(0)}$ and $a_{\mu x}^{(0)}$, where both $1/N$ quantum corrections (self-energies) given by $\varphi_{x}^{(1)}$ and $a_{\mu x}^{(1)}$ and vertex corrections described by the renormalized vierbein tensor $e_{a x}^{\nu (1)}$ have been introduced self-consistently through the renormalization group transformation with background dual fields.

\subsubsection{Recursive renormalization group transformations}

The above expression of the partition function suggests the following recursive form after the $f^{th}$ renormalization group transformation
\bqa && Z[\phi_{\alpha x}, A_{B \mu x}] = \int D \psi_{\alpha x} \Pi_{k = 0}^{f} D \varphi_{x}^{(k)} D a_{\mu x}^{(k)} D e_{a x}^{\mu (k)} D \theta^{a (k)}_{\mu x} D p_{x}^{(k)} D q_{x}^{(k)} D y_{x}^{(k)} D w_{x}^{(k)} \nn && \exp\Bigg[ - \int d^{D} x \sqrt{g_{x}^{(f)}} \Bigg\{ \bar{\psi}_{\alpha x} \gamma^{a} e_{a x}^{\mu (f)} \Big( \partial_{\mu} - i q a_{\mu x}^{(f)} - \frac{i}{4} \omega_{\mu x}^{a'b' (f)} \sigma_{a'b'} \Big) \psi_{\alpha x} + (m - i \varphi_{x}^{(f)}) \bar{\psi}_{\alpha x} \psi_{\alpha x} \nn && + \bar{\phi}_{\alpha x} p_{x}^{(f) \dagger} \psi_{\alpha x} + \bar{\psi}_{\alpha x} p_{x}^{(f)} \phi_{\alpha x} + \bar{\phi}_{\alpha x} y_{x}^{(f)} \phi_{\alpha x} \Bigg\} - N \int d^{D} x \sqrt{g_{x}^{(0)}} \Bigg\{ \frac{1}{2 \lambda_{\chi}} \varphi_{x}^{(0) 2} + \frac{q}{2 \lambda_{j}} g_{x}^{\mu\nu (0)} (a_{\mu x}^{(0)} - A_{B \mu x}) (a_{\nu x}^{(0)} - A_{B \nu x}) \nn && - \theta_{\mu x}^{a (0)} (e_{a x}^{\mu (0)} - e_{B a x}^{\mu}) - q_{x}^{(0) \dagger} (p_{x}^{(0)} - p_{B x}) - H.c. - w_{x}^{(0) \dagger} (y_{x}^{(0)} - y_{B x}) - H.c. \Bigg\} \nn && - N (2 d z) \sum_{k = 1}^{f} \int d^{D} x \sqrt{g_{x}^{(k-1)}} \Bigg\{ \frac{1}{2 \lambda_{\chi}} \Big(\frac{\varphi_{x}^{(k)} - \varphi_{x}^{(k-1)}}{2 d z}\Big)^{2} + \frac{\mathcal{C}_{\varphi}}{2} g_{x}^{\mu\nu (k-1)} (\partial_{\mu} \varphi_{x}^{(k)}) (\partial_{\nu} \varphi_{x}^{(k)}) + \mathcal{C}_{\xi} R_{x}^{(k-1)} \varphi_{x}^{(k) 2} \nn && + \mathcal{C}_{mix} F_{x}^{\mu\nu (k)} (\partial_{\mu} \varphi_{x}^{(k)}) (\partial_{\nu} \varphi_{x}^{(k)}) + \frac{q}{2 \lambda_{j}} g_{x}^{\mu\nu (k-1)} \Bigg(\frac{a_{\mu x}^{(k)} - a_{\mu x}^{(k-1)}}{2 d z} + \frac{1}{2 d z D} e^{a (k-1)}_{\mu x} \mbox{tr} \Big[ \gamma_{a} \Big\{ \gamma^{c} e_{c x}^{\lambda (k-1)} \Big( \partial_{\lambda} - i q a_{\lambda x}^{(k)} \nn && - \frac{i}{4} \omega_{\lambda x}^{c'd' (k-1)} \sigma_{c'd'} \Big) G_{xx'}^{(k-1)}[e_{a x}^{\lambda (k-1)},a_{\lambda x}^{(k)},\varphi_{x}^{(k)}] \Big\}_{x' \rightarrow x} \gamma^{b} \Big] e_{b x}^{\kappa (k-1)} a_{\kappa x}^{(k)}\Bigg) \Bigg(\frac{a_{\nu x}^{(k)} - a_{\nu x}^{(k-1)}}{2 d z} \nn && + \frac{1}{2 d z D} e^{w (k-1)}_{\nu x} \mbox{tr} \Big[ \gamma_{w} \Big\{ \gamma^{g} e_{g x}^{\eta (k-1)} \Big( \partial_{\eta} - i q a_{\eta x}^{(k)} - \frac{i}{4} \omega_{\eta x}^{g'h' (k-1)} \sigma_{g'h'} \Big) G_{xx'}^{(k-1)}[e_{a x}^{\eta (k-1)},a_{\eta x}^{(k)},\varphi_{x}^{(k)}] \Big\}_{x' \rightarrow x} \gamma^{l} \Big] e_{l x}^{\epsilon (k-1)} a_{\epsilon x}^{(k)}\Bigg) \nn && + \mathcal{C}_{F} F_{\mu\nu x}^{(k)} F_{x}^{\mu\nu (k)} - \mathcal{C}_{\Lambda} + \mathcal{C}_{R} R_{x}^{(k-1)} \Bigg\} - N (2 d z) \sum_{k = 1}^{f} \int d^{D} x \sqrt{g_{x}^{(k)}} \Bigg\{ - \theta_{\mu x}^{a (k)} \Bigg( \frac{e_{a}^{\mu (k)} - e_{a}^{\mu (k-1)}}{2 d z} \nn && + \frac{1}{2 d z D} \mbox{tr} \Big[ \gamma_{a} \Big\{ \gamma^{c} e_{c x}^{\nu (k-1)} \Big( \partial_{\nu} - i q a_{\nu x}^{(k)} - \frac{i}{4} \omega_{\nu x}^{c'd' (k-1)} \sigma_{c'd'} \Big) G_{xx'}^{(k-1)}[e_{a x}^{\nu (k-1)},a_{\nu x}^{(k)},\varphi_{x}^{(k)}] \Big\}_{x' \rightarrow x} \gamma^{b} \Big] e_{b}^{\mu (k-1)} \Bigg) \nn && - q_{x}^{(k) \dagger} \Bigg( \frac{p_{x}^{(k)} - p_{x}^{(k-1)}}{2 d z} + \frac{1}{2 d z} \Big\{ \gamma^{c} e_{c x}^{\mu (k-1)} \Big( \partial_{\mu} - i q a_{\mu x}^{(k)} - \frac{i}{4} \omega_{\mu x}^{c'd' (k-1)} \sigma_{c'd'} \Big) G_{xx'}^{(k-1)}[e_{a x}^{\nu (k-1)},a_{\nu x}^{(k)},\varphi_{x}^{(k)}] \Big\}_{x' \rightarrow x} p_{x}^{(k-1)} \Bigg) \nn && - H.c. - w_{x}^{(k) \dagger} \Bigg( \frac{y_{x}^{(k)} - y_{x}^{(k-1)}}{2 d z} + \frac{1}{2 d z} p_{x}^{(k-1) \dagger} \Big( G_{xx'}^{(k-1)}[e_{a x}^{\nu (k-1)},a_{\nu x}^{(k)},\varphi_{x}^{(k)}] \Big)_{x' \rightarrow x} p_{x}^{(k-1)} \Bigg) - H.c. \Bigg\} \Bigg] , \label{Recursive_RG_discrete} \eqa
where the gauge choice
\bqa && \alpha^{(k)} = 1 , ~~~~~ \beta^{(k)} = 1 , ~~~~~ \delta^{(k)} = 1 \eqa
has been taken into account. The Green's function is
\bqa && \Big\{ \gamma^{a} e_{a x}^{\nu (k-1)} \Big( \partial_{\nu} - i q a_{\nu x}^{(k)} - \frac{i}{4} \omega_{\nu x}^{a'b' (k-1)} \sigma_{a'b'} \Big) + \frac{1}{ 2 d z } (m - i \varphi_{x}^{(k)}) \Big\} G_{xx'}^{(k-1)}[e_{a x}^{\nu (k-1)},a_{\nu x}^{(k)},\varphi_{x}^{(k)}] = \frac{1}{\sqrt{g_{x}^{(k-1)}}} \delta^{(D)}(x-x') . \nn \eqa

\subsection{Effective geometric description for interacting Dirac fermions}

The final task is to rewrite the above partition function Eq. (\ref{Recursive_RG_discrete}) in the continuous coordinate representation instead of the discrete variable $k$. Resorting to
\bqa && (2 d z) \sum_{k = 1}^{f} \sqrt{g^{(k-1)}} \theta_{\mu x}^{a (k)} \Big( \frac{e_{a x}^{\mu (k)} - e_{a x}^{\mu (k-1)}}{2 d z} \Big) \Longrightarrow \int_{0}^{z_{f}} d z \sqrt{g(x,z)} ~ \theta_{\mu}^{a}(x,z) \partial_{z} e_{a}^{\mu}(x,z) \eqa
with $2 d z \sum_{k = 1}^{f} \Longrightarrow \int_{0}^{z_{f}} d z$ and $(e_{a x}^{\mu (k)} - e_{a x}^{\mu (k-1)}) / (2 d z) \Longrightarrow \partial_{z} e_{a}^{\mu}(x,z)$, we obtain
\bqa && Z[\phi_{\alpha}(x), A_{B \mu}(x)] = \int D \psi_{\alpha}(x) D \varphi(x,z) D a_{\mu}(x,z) D e_{a}^{\mu}(x,z) D p(x,z) D y(x,z) \nn && \delta\Big(e_{a}^{\mu}(x,0) - e_{B a}^{\mu}(x)\Big) \delta\Big(p(x,0) - p_{B}(x)\Big) \delta\Big(y(x,0) - y_{B}(x)\Big) \delta\Bigg( \partial_{z} e_{a}^{\mu}(x,z) + \frac{1}{\varepsilon D} \mbox{tr} \Big[ \gamma_{a} \Big\{ \gamma^{c} e_{c}^{\nu}(x,z) \Big( \partial_{\nu} \nn && - i q a_{\nu}(x,z) - \frac{i}{4} \omega_{\nu}^{c'd'}(x,z) \sigma_{c'd'} \Big) G_{xx'}[e_{a}^{\nu}(x,z),a_{\nu}(x,z),\varphi(x,z)] \Big\}_{x' \rightarrow x} \gamma^{b} \Big] e_{b}^{\mu}(x,z) \Bigg) \delta\Bigg( \partial_{z} p(x,z) \nn && + \frac{1}{\varepsilon} \Big\{ \gamma^{c} e_{c}^{\mu}(x,z) \Big( \partial_{\mu} - i q a_{\mu}(x,z) - \frac{i}{4} \omega_{\mu}^{c'd'}(x,z) \sigma_{c'd'} \Big) G_{xx'}[e_{a}^{\nu}(x,z),a_{\nu}(x,z),\varphi(x,z)] \Big\}_{x' \rightarrow x} p(x,z) \Bigg) \nn && \delta\Bigg( \partial_{z} y(x,z) + \frac{1}{\varepsilon} p^{\dagger}(x,z) \Big( G_{xx'}[e_{a}^{\nu}(x,z),a_{\nu}(x,z),\varphi(x,z)] \Big)_{x' \rightarrow x} p(x,z) \Bigg) \nn && \exp\Bigg[ - \int d^{D} x \sqrt{g(x,z_{f})} \Bigg\{ \bar{\psi}_{\alpha}(x) \gamma^{a} e_{a}^{\mu}(x,z_{f}) \Big( \partial_{\mu} - i q a_{\mu}(x,z_{f}) - \frac{i}{4} \omega_{\mu}^{a'b'}(x,z_{f}) \sigma_{a'b'} \Big) \psi_{\alpha}(x) \nn && + [m - i \varphi(x,z_{f})] \bar{\psi}_{\alpha}(x) \psi_{\alpha}(x) + \bar{\phi}_{\alpha}(x) p^{\dagger}(x,z_{f}) \psi_{\alpha}(x) + \bar{\psi}_{\alpha}(x) p(x,z_{f}) \phi_{\alpha}(x) + \bar{\phi}_{\alpha}(x) y(x,z_{f}) \phi_{\alpha}(x) \Bigg\} \nn && - N \int d^{D} x \sqrt{g(x,0)} \Bigg\{ \frac{1}{2 \lambda_{\chi}} [\varphi(x,0)]^{2} + \frac{q}{2 \lambda_{j}} g^{\mu\nu}(x,0) [a_{\mu}(x,0) - A_{B \mu}(x)] [a_{\nu}(x,0) - A_{B \nu}(x)] \Bigg\} \nn && - N \int_{0}^{z_{f}} d z \int d^{D} x \sqrt{g(x,z)} \Bigg\{ \frac{1}{2 \lambda_{\chi}} [\partial_{z} \varphi(x,z)]^{2} + \frac{\mathcal{C}_{\varphi}}{2} g^{\mu\nu}(x,z) [\partial_{\mu} \varphi(x,z)] [\partial_{\nu} \varphi(x,z)] + \mathcal{C}_{\xi} R(x,z) [\varphi(x,z)]^{2} \nn && + \mathcal{C}_{mix} F^{\mu\nu}(x,z) [\partial_{\mu} \varphi(x,z)] [\partial_{\nu} \varphi(x,z)] + \frac{q}{2 \lambda_{j}} g^{\mu\nu}(x,z) \Bigg(\partial_{z} a_{\mu} (x,z) + \frac{1}{\varepsilon D} e^{a}_{\mu}(x,z) \mbox{tr} \Big[ \gamma_{a} \Big\{ \gamma^{c} e_{c}^{\lambda}(x,z) \Big( \partial_{\lambda} \nn && - i q a_{\lambda}(x,z) - \frac{i}{4} \omega_{\lambda}^{c'd'}(x,z) \sigma_{c'd'} \Big) G_{xx'}[e_{a}^{\lambda}(x,z),a_{\lambda}(x,z),\varphi(x,z)] \Big\}_{x' \rightarrow x} \gamma^{b} \Big] e_{b}^{\kappa}(x,z) a_{\kappa}(x,z)\Bigg) \Bigg(\partial_{z} a_{\nu}(x,z) \nn && + \frac{1}{\varepsilon D} e^{w}_{\nu}(x,z) \mbox{tr} \Big[ \gamma_{w} \Big\{ \gamma^{g} e_{g}^{\eta}(x,z) \Big( \partial_{\eta} - i q a_{\eta}(x,z) - \frac{i}{4} \omega_{\eta}^{g'h'}(x,z) \sigma_{g'h'} \Big) G_{xx'}[e_{a}^{\eta}(x,z),a_{\eta}(x,z),\varphi(x,z)] \Big\}_{x' \rightarrow x} \gamma^{l} \Big] \nn && \times e_{l}^{\epsilon}(x,z) a_{\epsilon}(x,z)\Bigg) + \frac{\mathcal{C}}{4} F_{\mu\nu}(x,z) F^{\mu\nu}(x,z) + \frac{1}{2 \kappa} \Big( R(x,z) - 2 \Lambda \Big) \Bigg\} \Bigg] . \eqa
Here, we introduced changes in notations as
\bqa && \mathcal{C}_{R} = \frac{1}{2 \kappa} , ~~~~~ \frac{\mathcal{C}_{\Lambda}}{\mathcal{C}_{R}} = 2 \Lambda , ~~~~~ \mathcal{C}_{F} = \frac{\mathcal{C}}{4} \eqa
and
\bqa && \varepsilon \equiv 2 d z . \eqa
The Green's function of high-energy Dirac fermions is given by
\bqa && \Big\{ \gamma^{a} e_{a}^{\nu}(x,z) \Big( \partial_{\nu} - i q a_{\nu}(x,z) - \frac{i}{4} \omega_{\nu}^{a'b'}(x,z) \sigma_{a'b'} \Big) + \frac{m - i \varphi(x,z)}{ \varepsilon } \Big\} G_{xx'}[e_{a}^{\nu}(x,z),a_{\nu}(x,z),\varphi(x,z)] \nn && = \frac{1}{\sqrt{g(x,z)}} \delta^{(D)}(x-x') . \eqa
$z_{f} = f d z$ defines the size of an extra-dimensional space, where $f$ is the number of iteration steps of recursive renormalization group transformations and $d z$ is the renormalization group scale discussed before. $z_{f}$ is proportional to a UV cutoff, which defines an effective field theory of interacting Dirac fermions.

Physics of this partition function is clear. We recall an effective UV action given by
\bqa && \mathcal{S}_{UV} = \int d^{D} x \sqrt{g(x,0)} \Big\{ \bar{\psi}_{\alpha}(x) \gamma^{a} e_{a}^{\mu}(x,0) \Big( \partial_{\mu} - i q a_{\mu}(x,0) - \frac{i}{4} \omega_{\mu}^{a'b'}(x,0) \sigma_{a'b'} \Big) \psi_{\alpha}(x) + [m - i \varphi(x,0)] \bar{\psi}_{\alpha}(x) \psi_{\alpha}(x) \nn && + [\bar{\phi}_{\alpha}(x) p^{\dagger}(x,0) \psi_{\alpha}(x) + \bar{\psi}_{\alpha}(x) p(x,0) \phi_{\alpha}(x)] + \frac{N}{2 \lambda_{\chi}} [\varphi(x,0)]^{2} + \frac{N q}{2 \lambda_{j}} g^{\mu\nu}(x,0) [a_{\mu}(x,0) - A_{B \mu}(x)] [a_{\nu}(x,0) - A_{B \nu}(x)] \Big\} . \nonumber \eqa
%
%
On the other hand, the IR effective action is given by \bqa && \mathcal{S}_{IR} = \int d^{D} x \sqrt{g(x,z_{f})} \Big\{ \bar{\psi}_{\alpha}(x) \gamma^{a} e_{a}^{\mu}(x,z_{f}) \Big( \partial_{\mu} - i q a_{\mu}(x,z_{f}) - \frac{i}{4} \omega_{\mu}^{a'b'}(x,z_{f}) \sigma_{a'b'} \Big) \psi_{\alpha}(x) \nn && + [m - i \varphi(x,z_{f})] \bar{\psi}_{\alpha}(x) \psi_{\alpha}(x) + \bar{\phi}_{\alpha}(x) p^{\dagger}(x,z_{f}) \psi_{\alpha}(x) + \bar{\psi}_{\alpha}(x) p(x,z_{f}) \phi_{\alpha}(x) + \bar{\phi}_{\alpha}(x) y(x,z_{f}) \phi_{\alpha}(x) \Big\} , \nonumber \eqa
which takes into account all possible renormalizations such as field renormalization, mass renormalization, and interaction renormalization, described by the renormalized vierbein field $e_{a}^{\mu}(x,z_{f})$, the renormalized order-parameter field $\varphi(x,z_{f})$, and the renormalized U(1) gauge field $a_{\mu}(x,z_{f})$ with renormalized coefficients $p(x,z_{f})$ and $y(x,z_{f})$. The evolution equation for the vierbein field plays essentially the same role as renormalization group $\beta-$functions for coupling functions. On the other hand, the evolution equation of the U(1) gauge field (the order-parameter field) may be regarded as a Callan-Symanzik equation for the conserved U(1) current (non-conserved mass operator $N^{-1} \bar{\psi}_{\alpha}(x) \psi_{\alpha}(x)$) in the dual holographic description. In the large $N$ limit quantum fluctuations of all dynamical field variables are suppressed. As a result, we obtain an effective Maxwell equation for $a_{\mu}(x,z)$ and an effective Klein-Gordon equation for $\varphi(x,z)$ in an effective curved spacetime background given by \bqa && \partial_{z} e_{a}^{\mu}(x,z) = - \frac{1}{\varepsilon D} \mbox{tr} \Big[ \gamma_{a} \Big\{ \gamma^{c} e_{c}^{\nu}(x,z) \Big( \partial_{\nu} - i q a_{\nu}(x,z) - \frac{i}{4} \omega_{\nu}^{c'd'}(x,z) \sigma_{c'd'} \Big) G_{xx'}[e_{a}^{\nu}(x,z),a_{\nu}(x,z),\varphi(x,z)] \Big\}_{x' \rightarrow x} \gamma^{b} \Big] e_{b}^{\mu}(x,z) , \nonumber \eqa
which result from taking variations of the bulk effective action of the above partition function with respect to $a_{\mu}(x,z)$ and $\varphi(x,z)$, respectively. We emphasize that these coupled equations have nonlinearly intertwined structures, which introduced quantum corrections of such effective interactions into the partition function in the all-loop order \cite{Holographic_Liquid_Kim}. One may regard the dual order-parameter fields of $a_{\mu}(x,z)$ and $\varphi(x,z)$ as self-energy corrections and the emergent background geometry $e_{a}^{\mu}(x,z)$ as vertex corrections, both of which are taken into account self-consistently in this formulation. We point out that both the Maxwell type and Klein-Gordon type equations are given by the second-order differential equations for the extra-dimensional coordinate $z$. Two boundary conditions, referred to as UV and IR ones, are given by performing variations of UV and IR effective actions with respect to $a_{\mu}(x,0)$ ($\varphi(x,0)$) and $a_{\mu}(x,z_{f})$ ($\varphi(x,z_{f})$) in the large $N$ limit, which have to be supported by the linear-in-$z$ derivative UV and IR boundary terms resulting from the bulk effective action \cite{Gibbons_Hawking_York_I,Gibbons_Hawking_York_II}.

It is straightforward to find the IR Green's function in the large $N$ limit, given by the derivative of the IR effective action with respect to $\bar{\phi}_{\alpha}(x)$ and $\phi_{\alpha}(x')$. As a result, we obtain
\bqa && G(x,x') \equiv - \frac{1}{N} \Big\langle T_{\tau} \Big( \psi_{\alpha}(x) \bar{\psi}_{\alpha}(x') \Big) \Big\rangle \nn && = p^{\dagger}(x,z_{f}) \Big[ \Big\{ \gamma^{a} e_{a}^{\mu}(x,z_{f}) \Big( \partial_{\mu} - i q a_{\mu}(x,z_{f}) - \frac{i}{4} \omega_{\mu}^{a'b'}(x,z_{f}) \sigma_{a'b'} \Big) + m - i \varphi(x,z_{f}) \Big\}^{-1} \frac{1}{\sqrt{g(x,z_{f})}} \delta^{(D)}(x-x') \Big] p(x',z_{f}) \nn && + \frac{y(x,z_{f})}{\sqrt{g(x,z_{f})}} \delta^{(D)}(x-x') . \eqa
%
%
It is interesting to observe that two types of additional renormalizations appear. The UV-scale renormalization for the Green's function is given by $\frac{y(x,z_{f})}{\sqrt{g(x,z_{f})}} \delta^{(D)}(x-x')$ while the field renormalization is described by the renormalized coupling constant $p(x,z_{f})$. Their renormalization group flows are given by
\bqa && \partial_{z} p(x,z) = - \frac{1}{\varepsilon} \Big\{ \gamma^{c} e_{c}^{\mu}(x,z) \Big( \partial_{\mu} - i q a_{\mu}(x,z) - \frac{i}{4} \omega_{\mu}^{c'd'}(x,z) \sigma_{c'd'} \Big) G_{xx'}[e_{a}^{\nu}(x,z),a_{\nu}(x,z),\varphi(x,z)] \Big\}_{x' \rightarrow x} p(x,z) \nn \eqa
and
\bqa && \partial_{z} y(x,z) = - \frac{1}{\varepsilon} p^{\dagger}(x,z) \Big( G_{xx'}[e_{a}^{\nu}(x,z),a_{\nu}(x,z),\varphi(x,z)] \Big)_{x' \rightarrow x} p(x,z) , \eqa
respectively.

\subsection{Remarks on recursive renormalization group transformations}

Normally, to perform the renormalization group transformation, we would separate the slow and fast modes for each field and do the path integral for the fast modes to renormalize the dynamics of the slow modes. However, this does not look like what happens in this study. In this respect it is necessary to discuss how the same result could be obtained by a more standard approach. Recently, we applied momentum-space renormalization group transformations to the Kondo problem \cite{Holographic_Liquid_Kim}. Here, we separate slow and fast modes for each field in frequency space, and perform renormalization group transformations in a recursive way. As a result, we could describe the crossover regime successfully from the high-temperature decoupled local-moment fluctuating regime to the low-temperature Kondo-singlet local-Fermi-liquid state, where ``log" divergences are fully resummed through this recursive renormalization group framework \cite{Holographic_Liquid_Kim}. Unfortunately, we could not introduce the gravitational field to this momentum-space (or frequency-space) renormalization group transformation scheme, where the resulting effective action should be non-local in the reciprocal space, which prohibits us from applying the conventional renormalization group transformation scheme. However, it turns out that both the specific heat and the spin susceptibility for the impurity dynamics show reasonable match to those of the Bethe ansatz solution. This interesting result implies that the resulting (dual holographic) effective field theory takes into account quantum corrections in a non-perturbative way, i.e., in the all-loop order, although this does not mean that this solution is exact.

To figure out the present real-space renormalization group scheme more deeply, we applied the Kadanoff block-spin transformation to an effective field theory of the transverse-field Ising model in one spatial dimension, where an effective scalar field theory with their self-interactions has been considered \cite{Kadanoff_GR_Holography_Kim}. Here, we separate all field variables into those on even and odd sites and perform the path integral for even-site quantum fields. As a result, we could obtain an effective mean-field theory, where all the coupling functions such as effective mass and interaction vertices are fully renormalized and intertwined with the bulk dual order parameter field. In particular, comparing this effective mean-field theory with self-consistently renormalized interaction vertices (based on the Kadanoff block-spin transformation) to an effective holographic description of the scalar field theory (based on the present real-space renormalization group transformation), we could show that the renormalization group flow of the metric tensor in the latter renormalization group scheme is nothing but those of the coupling functions in the former renormalization group scheme \cite{Kadanoff_GR_Holography_Kim}. More precisely, we argued that these two partition functions are equivalent and represented the renormalization group flow of the metric tensor in terms of the renormalized mass parameter. In this respect we suspect that the present real-space renormalization group scheme may be a generalized version of the Kadanoff block-spin transformation in higher dimensions.

Furthermore, we investigated how to extract out renormalization coefficients such as field renormalization, mass renormalization, and interaction renormalization constants from the dual holographic effective field theory in a general ground. Reformulating the renormalized effective field theory with such renormalized coefficients in an effective holographic way, we could determine all these renormalization constants from the effective geometry in a non-perturbative way \cite{RG_GR_Holography_Kim}. It turns out that this proposal shares essentially the same spirit as the holographic renormalization group formulation for the renormalized effective on-shell action \cite{Holographic_Duality_IV,Holographic_Duality_V,Holographic_Duality_VI}.

The other point involved with the present recursive renormalization group transformation is that if the renormalization group transformation is performed in a conventional way, the resulting Wilsonian effective action should contain higher-derivative terms, suppressed by powers of the cutoff. It seems that this is not the result in the present framework.

An essential step in our recursive renormalization group transformation is the renormalization group transformation given by the path integral of heavy (large mass) fermion fields. As a result of this renormalization group transformation, both the metric tensor and the U(1) gauge field become truly renormalized. As discussed above and in our recent studies \cite{Holographic_Liquid_Kim,Kadanoff_GR_Holography_Kim}, the renormalization group flow of the metric tensor correspond to the renormalization group $\beta-$functions of the coupling functions. An interesting point is that the renormalization group flow of the metric tensor is given by the Green's function of high-energy fermions, which is the only dynamical information in this renormalization group transformation. If one sees this Green's function more carefully, it depends on not only the dual order parameter field, here the U(1) gauge field, but also the metric tensor. In other words, self-energy corrections given by the renormalization group flow of the dual order parameter field and vertex corrections given by that of the metric tensor are self-consistently and non-linearly intertwined, which takes into account renormalization effects non-perturbatively. Indeed, the IR boundary condition of the dual order parameter field corresponds to a mean-field equation of the order parameter field but with full renormalization effects given by the self-consistent renormalization group flow of the coupling functions \cite{Holographic_Liquid_Kim,Kadanoff_GR_Holography_Kim}. If one performs the gradient expansion with respect to the heavy mass for the dual order parameter field and the metric tensor in the Green's function, he/she would get higher-derivative terms, suppressed by higher powers of $d z$. In our reformulation, such terms have been resummed to be inside the Green's function.

Our final remark is on the origin of the diffeomorphism invariance of the emergent holographic description in the view of renormalization group transformations. First of all, the partition function itself is invariant under the spacetime dependent renormalization group transformation. Physically, the IR fixed-point physics should not depend on the spacetime dependent renormalization group transformation although the renormalization group path toward the IR fixed point does depend on the functional renormalization group scheme. This is because quantum fields of all energy scales have to be integrated out finally, regardless of the spacetime dependent speed of coarse-graining. This would be the origin of the diffeomorphism invariance in the bulk effective action. Here, it is broken by gauge fixing.

One subtlety is that there appear three kinds of local speeds of coarse-graining during the renormalization group transformation. First, we consider space-time independent constants for the speed of coarse-graining, given by the gauge fixing condition $\partial_{x} \alpha^{(0)} = \partial_{x} \beta^{(0)} = \partial_{x} \delta^{(0)} = 0$. Then, we obtain $\alpha^{(0)} = c_{1} \beta^{(0)}$ and $\delta^{(0)} = c_{2} \beta^{(0)}$, where $c_{1}$ and $c_{2}$ are positive numerical constants. Now, we introduce these speeds of coarse-graining into the renormalization group transformation for each field. These differences do not have any effects on the effective geometry since rescaling every field in an appropriate way, these differences can be erased. As a result, only one lapse function remains. Now, we allow the spacetime dependent speed of coarse-graining. In this case, we also have to introduce shift vectors into the effective action after the renormalization group transformation. At present, we do not understand how to construct such a diffeomorphism invariant effective action explicitly. Since we do not have such a formulation, we cannot answer whether these different spacetime-dependent speed-functions of coarse-graining can be gauged away by appropriate rescaling of quantum fields. How to implement the diffeomorphism invariance for the bulk effective action explicitly in the recursive renormalization-group transformation framework has to be investigated more carefully.

\section{Holographic dual effective field theory for interacting Dirac fermions}

\subsection{Introduction of effective tensor-type interactions}

Although the renormalization group flow of the vierbein tensor originates from vacuum fluctuations of high-energy Dirac fermions, the vierbein tensor is not fully dynamical in contrast to the holographic duality conjecture \cite{Holographic_Duality_I,Holographic_Duality_II,Holographic_Duality_III, Holographic_Duality_IV,Holographic_Duality_V,Holographic_Duality_VI,Holographic_Duality_VII}. On the other hand, both the order-parameter field and U(1) gauge field are dynamical in the extra-dimensional space. We recall that the dynamics of these dual fields result from effective interactions between matter fields in the renormalization group transformation. In this respect we introduce effective tensor-field interactions into the partition function as follows \cite{TTbar_Deformation}
\bqa && Z[\phi_{\alpha}, A_{B \mu}] = \int D \psi_{\alpha} \exp\Big[ - \int d^{D} x \sqrt{g_{B}} \Big\{ \bar{\psi}_{\alpha} \gamma^{a} e_{B a}^{\mu} \Big( \partial_{\mu} - i q A_{B \mu} - \frac{i}{4} \omega_{B \mu}^{a'b'} \sigma_{a'b'} \Big) \psi_{\alpha} + m \bar{\psi}_{\alpha} \psi_{\alpha} \nn && + ( \bar{\phi}_{\alpha} p_{B}^{\dagger} \psi_{\alpha} + \bar{\psi}_{\alpha} p_{B} \phi_{\alpha}) + \frac{\lambda_{\chi}}{2 N} (\bar{\psi}_{\alpha} \psi_{\alpha})^{2} + \frac{q \lambda_{j}}{2 N} g^{\mu\nu}_{B} (\bar{\psi}_{\alpha} \gamma_{a} e_{B \mu}^{a} \psi_{\alpha}) (\bar{\psi}_{\beta} \gamma_{b} e_{B \nu}^{b} \psi_{\beta}) \nn && + \frac{\lambda_{t}}{2 N} g_{B}^{\mu\nu} \Big\{ \bar{\psi}_{\alpha} \gamma^{a} \Big( \partial_{\mu} - i q A_{B \mu} - \frac{i}{4} \omega_{B \mu}^{a'b'} \sigma_{a'b'} \Big) \psi_{\alpha} \Big\} \Big\{ \bar{\psi}_{\beta} \gamma^{a} \Big( \partial_{\nu} - i q A_{B \nu} - \frac{i}{4} \omega_{B \nu}^{c'd'} \sigma_{c'd'} \Big) \psi_{\beta} \Big\} \Big\} \Big] \label{Interaction_Dirac_Fermion_System} \eqa
It is natural to expect that the effective interaction term $\frac{\lambda_{t}}{2 N} g_{B}^{\mu\nu} \Big\{ \bar{\psi}_{\alpha} \gamma^{a} \Big( \partial_{\mu} - i q A_{B \mu} - \frac{i}{4} \omega_{B \mu}^{a'b'} \sigma_{a'b'} \Big) \psi_{\alpha} \Big\} \Big\{ \bar{\psi}_{\beta} \gamma^{a} \Big( \partial_{\nu} - i q A_{B \nu} - \frac{i}{4} \omega_{B \nu}^{c'd'} \sigma_{c'd'} \Big) \psi_{\beta} \Big\}$ is irrelevant in the renormalization group sense as long as the coupling constant $\lambda_{t}$ remains to be below a critical value. However, we point out that these tensor-type quantum fluctuations promote the emergent vierbein tensor to be fully dynamical \cite{TTbar_Deformation}. Although one can take into account an effective interaction term of the exact energy-momentum tensor in the effective Lagrangian, we mention that the above introduction of the effective interaction term is sufficient in discussing which approximation scheme has to be used for the vierbein renormalization-group transformation.

Performing the Hubbard-Stratonovich transformation for all types of effective interactions, we obtain
\bqa && Z[\phi_{\alpha}, A_{B \mu}] = \int D \psi_{\alpha} D \varphi^{(0)} D a_{\mu}^{(0)} D \tilde{e}_{a}^{\mu} D e_{a}^{\mu (0)} D \theta^{a (0)}_{\mu} D p^{(0)} D q^{(0)} \nn && \exp\Big[ - \int d^{D} x \sqrt{g^{(0)}} \Big\{ \bar{\psi}_{\alpha} \gamma^{a} (e_{a}^{\mu (0)} - i \tilde{e}_{a}^{\mu}) \Big( \partial_{\mu} - i q a_{\mu}^{(0)} - \frac{i}{4} \omega_{\mu}^{a'b' (0)} \sigma_{a'b'} \Big) \psi_{\alpha} + (m - i \varphi^{(0)}) \bar{\psi}_{\alpha} \psi_{\alpha} \nn && + ( \bar{\phi}_{\alpha} p_{B}^{\dagger} \psi_{\alpha} + \bar{\psi}_{\alpha} p_{B} \phi_{\alpha}) + \frac{N}{2 \lambda_{\chi}} \varphi^{(0) 2} + \frac{N q}{2 \lambda_{j}} g^{\mu\nu (0)} (a_{\mu}^{(0)} - A_{B \mu}) (a_{\nu}^{(0)} - A_{B \nu}) + \frac{N}{2 \lambda_{t}} g_{\mu\nu}^{(0)} \tilde{e}_{a}^{\mu} \tilde{e}_{a}^{\nu} \nn && - N \theta_{\mu}^{a (0)} (e_{a}^{\mu (0)} - e_{B a}^{\mu}) - N [q^{(0) \dagger} (p^{(0)} - p_{B}) + H.c.] \Big\} \Big] , \eqa
where $\tilde{e}_{a}^{\mu}$ has been introduced to decompose the tensor-type interaction term. Shifting the vierbein tensor in the following way
\bqa && e_{a}^{\mu (0)} \Longrightarrow e_{a}^{\mu (0)} + i \tilde{e}_{a}^{\mu} , \eqa
we obtain
\bqa && Z[\phi_{\alpha}, A_{B \mu}] = \int D \psi_{\alpha} D \varphi^{(0)} D a_{\mu}^{(0)} D \tilde{e}_{a}^{\mu} D e_{a}^{\mu (0)} D \theta^{a (0)}_{\mu} D p^{(0)} D q^{(0)} \nn && \exp\Big[ - \int d^{D} x \sqrt{g[e_{\mu}^{a (0)} + i \tilde{e}_{\mu}^{a}]} \Big\{ \bar{\psi}_{\alpha} \gamma^{a} e_{a}^{\mu (0)} \Big( \partial_{\mu} - i q a_{\mu}^{(0)} - \frac{i}{4} \omega_{\mu}^{a'b'}[e_{\mu}^{a (0)} + i \tilde{e}_{\mu}^{a}] \sigma_{a'b'} \Big) \psi_{\alpha} + (m - i \varphi^{(0)}) \bar{\psi}_{\alpha} \psi_{\alpha} \nn && + ( \bar{\phi}_{\alpha} p_{B}^{\dagger} \psi_{\alpha} + \bar{\psi}_{\alpha} p_{B} \phi_{\alpha}) + \frac{N}{2 \lambda_{\chi}} \varphi^{(0) 2} + \frac{N q}{2 \lambda_{j}} g^{\mu\nu}[e_{a}^{\mu (0)} + i \tilde{e}_{a}^{\mu}] (a_{\mu}^{(0)} - A_{B \mu}) (a_{\nu}^{(0)} - A_{B \nu}) + \frac{N}{2 \lambda_{t}} g_{\mu\nu}[e_{\mu}^{a (0)} + i \tilde{e}_{\mu}^{a}] \tilde{e}_{a}^{\mu} \tilde{e}_{a}^{\nu} \nn && - N \theta_{\mu}^{a (0)} (e_{a}^{\mu (0)} + i \tilde{e}_{a}^{\mu} - e_{B a}^{\mu}) - N [ q^{(0) \dagger} (p^{(0)} - p_{B}) + H.c.] \Big\} \Big] . \eqa
Here, our object is to perform the path integral $\int D \theta_{\mu}^{a (0)} D \tilde{e}_{a}^{\mu}$. The approximation that we have to use is to keep quantum fluctuations of the vierbein tensor up to the linear order. In other words, we neglect the $i \tilde{e}_{a}^{\mu}$ contribution in the determinant $\sqrt{g[e_{\mu}^{a (0)} + i \tilde{e}_{\mu}^{a}]}$, spin connection $\omega_{\mu}^{a'b'}[e_{\mu}^{a (0)} + i \tilde{e}_{\mu}^{a}]$, and metric tensor $g_{\mu\nu}[e_{\mu}^{a (0)} + i \tilde{e}_{\mu}^{a}]$, given by the following partition function
\bqa && Z[\phi_{\alpha}, A_{B \mu}] \approx \int D \psi_{\alpha} D \varphi^{(0)} D a_{\mu}^{(0)} D \tilde{e}_{a}^{\mu} D e_{a}^{\mu (0)} D \theta^{a (0)}_{\mu} D p^{(0)} D q^{(0)} \nn && \exp\Big[ - \int d^{D} x \sqrt{g^{(0)}} \Big\{ \bar{\psi}_{\alpha} \gamma^{a} e_{a}^{\mu (0)} \Big( \partial_{\mu} - i q a_{\mu}^{(0)} - \frac{i}{4} \omega_{\mu}^{a'b' (0)} \sigma_{a'b'} \Big) \psi_{\alpha} + (m - i \varphi^{(0)}) \bar{\psi}_{\alpha} \psi_{\alpha} \nn &&+ ( \bar{\phi}_{\alpha} p_{B}^{\dagger} \psi_{\alpha} + \bar{\psi}_{\alpha} p_{B} \phi_{\alpha}) + \frac{N}{2 \lambda_{\chi}} \varphi^{(0) 2} + \frac{N q}{2 \lambda_{j}} g^{\mu\nu (0)} (a_{\mu}^{(0)} - A_{B \mu}) (a_{\nu}^{(0)} - A_{B \nu}) + \frac{N}{2 \lambda_{t}} g_{\mu\nu}^{(0)} \tilde{e}_{a}^{\mu} \tilde{e}_{a}^{\nu} \nn && - N \theta_{\mu}^{a (0)} (e_{a}^{\mu (0)} + i \tilde{e}_{a}^{\mu} - e_{B a}^{\mu}) - N [ q^{(0) \dagger} (p^{(0)} - p_{B}) + H.c.] \Big\} \Big] . \eqa

Now, it is straightforward to perform the path integral $\int D \theta_{\mu}^{a (0)} D \tilde{e}_{a}^{\mu}$. As a result, we obtain
\bqa && Z[\phi_{\alpha}, A_{B \mu}] \approx \int D \psi_{\alpha} D \varphi^{(0)} D a_{\mu}^{(0)} D e_{a}^{\mu (0)} D p^{(0)} D q^{(0)} \nn && \exp\Big[ - \int d^{D} x \sqrt{g^{(0)}} \Big\{ \bar{\psi}_{\alpha} \gamma^{a} e_{a}^{\mu (0)} \Big( \partial_{\mu} - i q a_{\mu}^{(0)} - \frac{i}{4} \omega_{\mu}^{a'b' (0)} \sigma_{a'b'} \Big) \psi_{\alpha} + (m - i \varphi^{(0)}) \bar{\psi}_{\alpha} \psi_{\alpha} + ( \bar{\phi}_{\alpha} p_{B}^{\dagger} \psi_{\alpha} + \bar{\psi}_{\alpha} p_{B} \phi_{\alpha}) \nn && + \frac{N}{2 \lambda_{\chi}} \varphi^{(0) 2} + \frac{N q}{2 \lambda_{j}} g^{\mu\nu (0)} (a_{\mu}^{(0)} - A_{B \mu}) (a_{\nu}^{(0)} - A_{B \nu}) - \frac{N}{2 \lambda_{t}} g_{\mu\nu}^{(0)} (e_{a}^{\mu (0)} - e_{B a}^{\mu}) (e_{a}^{\nu (0)} - e_{B a}^{\nu}) \nn && - N [q^{(0) \dagger} (p^{(0)} - p_{B}) + H.c.] \Big\} \Big] , \eqa
where the Gaussian-fluctuation term has been generated for the vierbein tensor as both the order parameter and U(1) gauge field.

One can repeat essentially the same renormalization group transformation for the vierbein tensor as that of both the order parameter and U(1) gauge field. Introducing an auxiliary vierbein field into the effective action, separating both vierbein fields into slow and fast degrees of freedom as those of collective dual field variables, and doing the path integral with respect to the heavy vierbein field, one can implement the renormalization group transformation for the vierbein field. Here, an essential point of this renormalization group transformation different from that for both the order parameter and U(1) gauge field is that we have to resort to the linear approximation for the path integral of the heavy vierbein field, where nonlinear contributions give rise to various complicated terms beyond our control. Neglecting such nonlinear contributions and performing the path integral for the vierbein tensor at the Gaussian level, we obtain an effective interaction term between tensor fields, given by essentially the same form of the original effective interaction but with a small coefficient proportional to $d z$. Performing the Hubbard-Stratonovich transformation for this newly generated effective interaction term, we obtain an effective renormalized action for the vierbein tensor as that of both the order parameter and U(1) gauge field. Together with the renormalization group transformation for the Dirac spinor field, we update both the vierbein tensor and the U(1) gauge field and find an effective renormalized action after the first renormalization group transformation. Finally, the resulting recursive formula in the partition function is translated into
\bqa && Z[A_{B \mu}(x)] = \int D \psi_{\alpha}(x) D \varphi(x,z) D a_{\mu}(x,z) D e_{a}^{\mu}(x,z) \exp\Bigg[ - \int d^{D} x \sqrt{g(x,z_{f})} \Bigg\{ \bar{\psi}_{\alpha}(x) \gamma^{a} e_{a}^{\mu}(x,z_{f}) \Big( \partial_{\mu} \nn && - i q a_{\mu}(x,z_{f}) - \frac{i}{4} \omega_{\mu}^{a'b'}(x,z_{f}) \sigma_{a'b'} \Big) \psi_{\alpha}(x) + [m - i \varphi(x,z_{f})] \bar{\psi}_{\alpha}(x) \psi_{\alpha}(x) \Bigg\} - N \int d^{D} x \sqrt{g(x,0)} \Bigg\{ \frac{1}{2 \lambda_{\chi}} [\varphi(x,0)]^{2} \nn && + \frac{q}{2 \lambda_{j}} g^{\mu\nu}(x,0) [a_{\mu}(x,0) - A_{B \mu}(x)] [a_{\nu}(x,0) - A_{B \nu}(x)] - \frac{N}{2 \lambda_{t}} g_{\mu\nu}(x,0) [e_{a}^{\mu}(x,0) - e_{B a}^{\mu}(x)] [e_{a}^{\nu}(x,0) - e_{B a}^{\nu}(x)] \Bigg\} \nn && - N \int_{0}^{z_{f}} d z \int d^{D} x \sqrt{g(x,z)} \Bigg\{ \frac{1}{2 \lambda_{\chi}} [\partial_{z} \varphi(x,z)]^{2} + \frac{\mathcal{C}_{\varphi}}{2} g^{\mu\nu}(x,z) [\partial_{\mu} \varphi(x,z)] [\partial_{\nu} \varphi(x,z)] \nn && + \mathcal{C}_{\xi} R(x,z) [\varphi(x,z)]^{2} + \mathcal{C}_{mix} F^{\mu\nu}(x,z) [\partial_{\mu} \varphi(x,z)] [\partial_{\nu} \varphi(x,z)] \Bigg\} \nn && - N \int_{0}^{z_{f}} d z \int d^{D} x \sqrt{g(x,z)} \Bigg\{ \frac{q}{2 \lambda_{j}} g^{\mu\nu}(x,z) \Bigg(\partial_{z} a_{\mu} (x,z) + \frac{1}{\varepsilon D} e^{a}_{\mu}(x,z) \mbox{tr} \Big[ \gamma_{a} \Big\{ \gamma^{c} e_{c}^{\lambda}(x,z) \Big( \partial_{\lambda} \nn && - i q a_{\lambda}(x,z) - \frac{i}{4} \omega_{\lambda}^{c'd'}(x,z) \sigma_{c'd'} \Big) G_{xx'}[e_{a}^{\lambda}(x,z),a_{\lambda}(x,z),\varphi(x,z)] \Big\}_{x' \rightarrow x} \gamma^{b} \Big] e_{b}^{\kappa}(x,z) a_{\kappa}(x,z)\Bigg) \Bigg(\partial_{z} a_{\nu}(x,z) \nn && + \frac{1}{\varepsilon D} e^{w}_{\nu}(x,z) \mbox{tr} \Big[ \gamma_{w} \Big\{ \gamma^{g} e_{g}^{\eta}(x,z) \Big( \partial_{\eta} - i q a_{\eta}(x,z) - \frac{i}{4} \omega_{\eta}^{g'h'}(x,z) \sigma_{g'h'} \Big) G_{xx'}[e_{a}^{\eta}(x,z),a_{\eta}(x,z),\varphi(x,z)] \Big\}_{x' \rightarrow x} \gamma^{l} \Big] \nn && \times e_{l}^{\epsilon}(x,z) a_{\epsilon}(x,z)\Bigg) + \frac{\mathcal{C}}{4} F_{\mu\nu}(x,z) F^{\mu\nu}(x,z) \Bigg\} \nn && - N \int_{0}^{z_{f}} d z \int d^{D} x \sqrt{g(x,z)} \Bigg\{ - \frac{N}{2 \lambda_{t}} g_{\mu\nu}(x,z) \Bigg( \partial_{z} e_{a}^{\mu}(x,z) + \frac{1}{\varepsilon D} \mbox{tr} \Big[ \gamma_{a} \Big\{ \gamma^{c} e_{c}^{\lambda}(x,z) \Big( \partial_{\lambda} - i q a_{\lambda}(x,z) \nn && - \frac{i}{4} \omega_{\lambda}^{c'd'}(x,z) \sigma_{c'd'} \Big) G_{xx'}[e_{c}^{\lambda}(x,z),a_{\lambda}(x,z),\varphi(x,z)] \Big\}_{x' \rightarrow x} \gamma^{b} \Big] e_{b}^{\mu}(x,z) \Bigg) \Bigg( \partial_{z} e_{a}^{\nu}(x,z) + \frac{1}{\varepsilon D} \mbox{tr} \Big[ \gamma_{a} \Big\{ \gamma^{g} e_{g}^{\eta}(x,z) \Big( \partial_{\eta} \nn && - i q a_{\eta}(x,z) - \frac{i}{4} \omega_{\eta}^{g'h'}(x,z) \sigma_{g'h'} \Big) G_{xx'}[e_{g}^{\eta}(x,z),a_{\eta}(x,z),\varphi(x,z)] \Big\}_{x' \rightarrow x} \gamma^{l} \Big] e_{l}^{\nu}(x,z) \Bigg) + \frac{1}{2 \kappa} \Big( R(x,z) - 2 \Lambda \Big) \Bigg\} \Bigg] , ~~~~~ \eqa
where an extra-dimensional space has been introduced with a coordinate $z$ as before. In addition, the source terms introduced before for the IR Green's function have been omitted for the simplicity of presentation. Although most terms remain same as before, the gravity sector is now fully dynamical in the extra-dimensional space, which originates from effective interactions between energy-momentum tensors. More generally speaking, one may say that global symmetries are uplifted into local ones, consistent with the holographic duality conjecture, even if this bulk effective action is a gauge-fixed version and those gauge symmetries are explicitly broken by gauge fixing. The Green's function of the heavy Dirac spinor is given by
\bqa && \Big\{ \gamma^{a} e_{a}^{\nu}(x,z) \Big( \partial_{\nu} - i q a_{\nu}(x,z) - \frac{i}{4} \omega_{\nu}^{a'b'}(x,z) \sigma_{a'b'} \Big) + \frac{m - i \varphi(x,z)}{ \varepsilon } \Big\} G_{xx'}[e_{a}^{\nu}(x,z),a_{\nu}(x,z),\varphi(x,z)] \nn && = \frac{1}{\sqrt{g(x,z)}} \delta^{(D)}(x-x') . \nonumber \eqa

We emphasize that the renormalization group flow of the vierbein tensor is highly nonlinear and intertwined with the renormalization group flow of both the order-parameter field and the U(1) gauge field in the extra-dimensional space although we resort to the linear approximation for the renormalization group transformation of the vierbein tensor. The origin of this nonlinearly intertwined renormalization group flow is in the feedback effect of the Green's function, where the Green's function is determined by all these field variables.

To compare this holographic dual effective field theory with that of the holographic duality conjecture, we perform the path integral with respect to the Dirac spinor field. As a result, we obtain the following expression of the partition function
\bqa && Z[A_{B \mu}(x)] = \int D \varphi(x,z) D a_{\mu}(x,z) D e_{a}^{\mu}(x,z) \exp\Big\{ - \mathcal{S}_{Bulk}[\varphi(x,z), a_{\mu}(x,z), e_{a}^{\mu}(x,z)] \nn && - \mathcal{S}_{IR}[\varphi(x,z_{f}), a_{\mu}(x,z_{f}), e_{a}^{\mu}(x,z_{f})] - \mathcal{S}_{UV}[\varphi(x,0), a_{\mu}(x,0), e_{a}^{\mu}(x,0);A_{B \mu}(x)] \Big\} . \eqa
The IR effective action is
%
%
\bqa && \mathcal{S}_{IR}[\varphi(x,z_{f}), a_{\mu}(x,z_{f}), e_{a}^{\mu}(x,z_{f})] = N \int d^{D} x \sqrt{g(x,z_{f})} \Bigg\{ \frac{\mathcal{C}_{\varphi}^{f}}{2} g^{\mu\nu}(x,z_{f}) [\partial_{\mu} \varphi(x,z_{f})] [\partial_{\nu} \varphi(x,z_{f})] \nn && + \mathcal{C}_{\xi}^{f} R(x,z_{f}) [\varphi(x,z_{f})]^{2} + \mathcal{C}_{mix}^{f} F^{\mu\nu}(x,z_{f}) [\partial_{\mu} \varphi(x,z_{f})] [\partial_{\nu} \varphi(x,z_{f})] + \frac{\mathcal{C}_{f}}{4} F_{\mu\nu}(x,z_{f}) F^{\mu\nu}(x,z_{f}) \nn && + \frac{1}{2 \kappa_{f}} \Big( R(x,z_{f}) - 2 \Lambda_{f} \Big) \Bigg\} , \eqa
which serves as IR boundary conditions for the order-parameter field $\varphi(x,z_{f})$, the U(1) gauge field $a_{\mu}(x,z_{f})$, and the vierbein tensor field or the metric tensor $g^{\mu\nu}(x,z_{f})$. All the positive coefficients $\mathcal{C}_{\varphi}^{f}$, $\mathcal{C}_{\xi}^{f}$, $\mathcal{C}_{mix}^{f}$, $\mathcal{C}_{f}$, $\kappa_{f}$, and $\Lambda_{f}$ are determined by the gradient expansion \cite{Gradient_Expansion_Gravity_I,Gradient_Expansion_Gravity_II} of \bqa && - N \mbox{tr}_{xx'} \ln \sqrt{g(x,z_{f})} \Big\{ \gamma^{a} e_{a}^{\mu}(x,z_{f}) \Big( \partial_{\mu} - i q a_{\mu}(x,z_{f}) - \frac{i}{4} \omega_{\mu}^{a'b'}(x,z_{f}) \sigma_{a'b'} \Big) + [m - i \varphi(x,z_{f})] \Big\} . \nonumber \eqa The bulk effective action is
\bqa && \mathcal{S}_{Bulk}[\varphi(x,z), a_{\mu}(x,z), e_{a}^{\mu}(x,z)] = N \int_{0}^{z_{f}} d z \int d^{D} x \sqrt{g(x,z)} \Bigg\{ \frac{1}{2 \lambda_{\chi}} [\partial_{z} \varphi(x,z)]^{2} + \frac{\mathcal{C}_{\varphi}}{2} g^{\mu\nu}(x,z) [\partial_{\mu} \varphi(x,z)] [\partial_{\nu} \varphi(x,z)] \nn && + \mathcal{C}_{\xi} R(x,z) [\varphi(x,z)]^{2} + \mathcal{C}_{mix} F^{\mu\nu}(x,z) [\partial_{\mu} \varphi(x,z)] [\partial_{\nu} \varphi(x,z)] \Bigg\} \nn && + N \int_{0}^{z_{f}} d z \int d^{D} x \sqrt{g(x,z)} \Bigg\{ \frac{q}{2 \lambda_{j}} g^{\mu\nu}(x,z) \Bigg(\partial_{z} a_{\mu} (x,z) + \frac{1}{\varepsilon D} e^{a}_{\mu}(x,z) \mbox{tr} \Big[ \gamma_{a} \Big\{ \gamma^{c} e_{c}^{\lambda}(x,z) \Big( \partial_{\lambda} - i q a_{\lambda}(x,z) \nn && - \frac{i}{4} \omega_{\lambda}^{c'd'}(x,z) \sigma_{c'd'} \Big) G_{xx'}[e_{a}^{\lambda}(x,z),a_{\lambda}(x,z),\varphi(x,z)] \Big\}_{x' \rightarrow x} \gamma^{b} \Big] e_{b}^{\kappa}(x,z) a_{\kappa}(x,z)\Bigg) \Bigg(\partial_{z} a_{\nu}(x,z) \nn && + \frac{1}{\varepsilon D} e^{w}_{\nu}(x,z) \mbox{tr} \Big[ \gamma_{w} \Big\{ \gamma^{g} e_{g}^{\eta}(x,z) \Big( \partial_{\eta} - i q a_{\eta}(x,z) - \frac{i}{4} \omega_{\eta}^{g'h'}(x,z) \sigma_{g'h'} \Big) G_{xx'}[e_{a}^{\eta}(x,z),a_{\eta}(x,z),\varphi(x,z)] \Big\}_{x' \rightarrow x} \gamma^{l} \Big] \nn && \times e_{l}^{\epsilon}(x,z) a_{\epsilon}(x,z)\Bigg) + \frac{\mathcal{C}}{4} F_{\mu\nu}(x,z) F^{\mu\nu}(x,z) \Bigg\} \nn && + N \int_{0}^{z_{f}} d z \int d^{D} x \sqrt{g(x,z)} \Bigg\{ - \frac{N}{2 \lambda_{t}} g_{\mu\nu}(x,z) \Bigg( \partial_{z} e_{a}^{\mu}(x,z) + \frac{1}{\varepsilon D} \mbox{tr} \Big[ \gamma_{a} \Big\{ \gamma^{c} e_{c}^{\lambda}(x,z) \Big( \partial_{\lambda} - i q a_{\lambda}(x,z) \nn && - \frac{i}{4} \omega_{\lambda}^{c'd'}(x,z) \sigma_{c'd'} \Big) G_{xx'}[e_{c}^{\lambda}(x,z),a_{\lambda}(x,z),\varphi(x,z)] \Big\}_{x' \rightarrow x} \gamma^{b} \Big] e_{b}^{\mu}(x,z) \Bigg) \Bigg( \partial_{z} e_{a}^{\nu}(x,z) + \frac{1}{\varepsilon D} \mbox{tr} \Big[ \gamma_{a} \Big\{ \gamma^{g} e_{g}^{\eta}(x,z) \Big( \partial_{\eta} \nn && - i q a_{\eta}(x,z) - \frac{i}{4} \omega_{\eta}^{g'h'}(x,z) \sigma_{g'h'} \Big) G_{xx'}[e_{g}^{\eta}(x,z),a_{\eta}(x,z),\varphi(x,z)] \Big\}_{x' \rightarrow x} \gamma^{l} \Big] e_{l}^{\nu}(x,z) \Bigg) + \frac{1}{2 \kappa} \Big( R(x,z) - 2 \Lambda \Big) \Bigg\} . \eqa
Compared to the IR effective action, there are Gaussian-fluctuation terms with $z-$derivatives for all dynamical fields. In addition, the coefficients of $\mathcal{C}_{\varphi}$, $\mathcal{C}_{\xi}$, $\mathcal{C}_{mix}$, $\mathcal{C}$, $\kappa$, and $\Lambda$ differ from those of $\mathcal{C}_{\varphi}^{f}$, $\mathcal{C}_{\xi}^{f}$, $\mathcal{C}_{mix}^{f}$, $\mathcal{C}_{f}$, $\kappa_{f}$, and $\Lambda_{f}$. To compensate these differences between coefficients, the IR boundary conditions given by the IR effective action have to be supported by linear-in-$z$ derivative terms of all dynamical fields at the IR boundary $z = z_{f}$, which appear from the bulk effective action by the method of integration-by-parts \cite{Gibbons_Hawking_York_I,Gibbons_Hawking_York_II}. The UV effective action is given by
\bqa && \mathcal{S}_{UV}[\varphi(x,0), a_{\mu}(x,0), e_{a}^{\mu}(x,0);A_{B \mu}(x)] = N \int d^{D} x \sqrt{g(x,0)} \Bigg\{ \frac{1}{2 \lambda_{\chi}} [\varphi(x,0)]^{2} \nn && + \frac{q}{2 \lambda_{j}} g^{\mu\nu}(x,0) [a_{\mu}(x,0) - A_{B \mu}(x)] [a_{\nu}(x,0) - A_{B \nu}(x)] - \frac{N}{2 \lambda_{t}} g_{\mu\nu}(x,0) [e_{a}^{\mu}(x,0) - e_{B a}^{\mu}(x)] [e_{a}^{\nu}(x,0) - e_{B a}^{\nu}(x)] \Bigg\} , \nn \eqa
which encodes initial values of all dynamical field variables, where the UV boundary conditions given by this UV effective action have to be also supported by linear-in-$z$ derivative terms at the UV boundary $z = 0$ resulting from the bulk effective action. This concludes our derivation for an effective dual holographic Einstein-Maxwell theory from the quantum field theory of interacting Dirac fermions.

We recall $g_{DD} = e_{D}^{a} e_{D}^{b} \delta_{ab} = 1$, given by the globally uniform speed of coarse graining, i.e., $\alpha(x,z) = \beta(x,z) = \delta(x,z) = 1$. We point out that there is an additional gauge choice, giving rise to $g_{\mu D} = e_{\mu}^{a} e_{D}^{b} \delta_{ab} = 0$ with $\mu = 0, ..., D-1$ in this effective bulk action. This gauge freedom originates from the invariance of the partition function with respect to $D-$dimensional diffeomorphism after the renormalization group transformation with $d z$, which means that the partition function has to be independent regardless of spacetime dependent UV cutoffs \cite{SungSik_Holography_I,SungSik_Holography_II,SungSik_Holography_III}. Furthermore, we have the gauge choice of $a_{D}(x,z) = 0$ in the Maxwell dynamics. A fully covariant formulation has been constructed in the absence of U(1) gauge fields, where $D-$dimensional Einstein-Hilbert action is uplifted into $(D+1)-$dimensional Einstein-Hilbert one via recursive renormalization group transformations \cite{SungSik_Holography_I,SungSik_Holography_II,SungSik_Holography_III}.

%
%

\subsection{Prescription}

The remaining task is to derive three types of coupled equations, referred to as effective Maxwell equations, effective Klein-Gordon equations, and effective Einstein equations, performing variations of the bulk effective action with respect to $a_{\mu}(x,z)$, $\varphi(x,z)$, and $e_{a}^{\mu}(x,z)$, respectively, and to solve them with both UV and IR boundary conditions, all of which are derived from this effective action functional in the large $N$ limit. Since we believe that this task has to be performed in a separate publication, here we give our prescription based on the holographic dual effective field theory discussed in this paper.

Considering how to obtain the effective hydrodynamics in the holographic duality conjecture \cite{Holographic_Liquid_Son_I,Holographic_Liquid_Son_II,Holographic_Liquid_Son_III, Holographic_Liquid_Son_IV}, we consider interacting Dirac fermions at finite temperatures, where an AdS black hole solution can arise \cite{Hawking_Page_Transition_I,Hawking_Page_Transition_II}. In this respect we separate effective geometry into time and space components. In addition, we consider small fluctuations of scalar fields, U(1) gauge fields, and vierbein fields, respectively, around their vacuum solution or the free-energy minimizing solution. In other words, we consider
\bqa && e_{a}^{\tau}(x,z) = \bar{e}_{a}^{\tau}(z) + \delta e_{a}^{\tau}(x,z) , ~~~~~ e_{a}^{i}(x,z) = \bar{e}_{a}^{i}(z) + \delta e_{a}^{i}(x,z) , \nn && a_{\tau}(x,z) = \bar{a}_{\tau}(z) + \delta a_{\tau}(x,z) , ~~~~~ a_{i}(x,z) = \bar{a}_{i}(z) + \delta a_{i}(x,z) , ~~~~~ \varphi(x,z) = \bar{\varphi}(z) + \delta \varphi(x,z) , \eqa
where the free-energy minimizing solution given by $\bar{e}_{a}^{\tau}(z)$, $\bar{e}_{a}^{i}(z)$, $\bar{a}_{\tau}(z)$, $\bar{a}_{i}(z)$, and $\bar{\varphi}(z)$ with $i = 1, ..., D - 1$ has been assumed to be translationally invariant in spacetime. Within this ansatz, the Green's function of the high-energy Dirac spinor field is given by
\bqa && G_{xx'}[\bar{e}_{a}^{\tau}(z) + \delta e_{a}^{\tau}(x,z),\bar{e}_{a}^{i}(z) + \delta e_{a}^{i}(x,z),\bar{a}_{\tau}(z) + \delta a_{\tau}(x,z),\bar{a}_{i}(z) + \delta a_{i}(x,z),\bar{\varphi}(z) + \delta \varphi(x,z)] \nn && = \bar{G}_{xx'}[\bar{e}_{a}^{\tau}(z),\bar{e}_{a}^{i}(z),\bar{a}_{\tau}(z),\bar{a}_{i}(z),\bar{\varphi}(z)] + \mathcal{G}_{xx'}^{n}[\bar{e}_{a}^{\tau}(z),\bar{e}_{a}^{i}(z),\bar{a}_{\tau}(z),\bar{a}_{i}(z),\bar{\varphi}(z)] \delta x_{n}(x,z) \eqa
with \bqa && \delta x_{n}(x,z) = [\delta x_{1}(x,z),\delta x_{2}(x,z),\delta x_{3}(x,z),\delta x_{4}(x,z),\delta x_{5}(x,z)] \equiv [\delta e_{a}^{\tau}(x,z),\delta e_{a}^{i}(x,z),\delta a_{\tau}(x,z),\delta a_{i}(x,z),\delta \varphi(x,z)] , \nonumber \eqa
which has been expanded up to the linear order in small fluctuations of all dynamical fields. Here, the mean-field propagator is
\bqa && \Big\{ \gamma^{a} \bar{e}_{a}^{\tau}(z) \Big( \partial_{\tau} - i q \bar{a}_{\tau}(z) \Big) + \gamma^{a} \bar{e}_{a}^{i}(z) \Big( \partial_{i} - i q \bar{a}_{i}(z) \Big) + \frac{m - i \bar{\varphi}(z)}{ \varepsilon } \Big\} \bar{G}_{xx'}[\bar{e}_{a}^{\tau}(z),\bar{e}_{a}^{i}(z),\bar{a}_{\tau}(z),\bar{a}_{i}(z),\bar{\varphi}(z)] \nn && = \frac{1}{\sqrt{\bar{g}(z)}} \delta^{(D)}(x-x') . \label{High_Energy_Green_Function_Mean_Field} \eqa

Introducing all these expressions into the effective action or the resulting coupled equations of motion and keeping all terms up to the linear order in these small fluctuations, we find two separate equations, which correspond to the mean-field equations at finite temperatures but with all-loop-order-resummed quantum corrections and linearized coupled equations of small fluctuations, respectively. Here, the first question is whether the non-perturbative mean-field equations at finite temperatures give an AdS black hole type solution above a certain critical or threshold temperature. Suppose we have the black hole solution as an effective background geometry. Then, the second question is whether the coupled linearized equations for small fluctuations of all dynamical fields give rise to diffusive normal modes for transverse fluctuations \cite{Holographic_Liquid_Son_I,Holographic_Liquid_Son_II,Holographic_Liquid_Son_III, Holographic_Liquid_Son_IV}. Expressing such weak perturbations of all dynamical fields in terms of external electromagnetic fields and temperature gradients, we obtain an effective on-shell action in terms of such external source fields. Taking derivatives for the effective on-shell action with respect to both external electromagnetic fields and temperature gradients, we find correlation functions of electromagnetic U(1) currents and energy-momentum tensors, which correspond to electrical, thermal, and thermoelectrical transport coefficients. This requires further investigations near future.

In this prescription, a key quantity which appears in all equations of motion for all dynamical fields is the frequency-momentum integral of the Green's function of high-energy Dirac fermions given by Eq. (\ref{High_Energy_Green_Function_Mean_Field}). It turns out that there appear UV divergences in this frequency-momentum integral of the Green's function when the UV cutoff is set to be infinite. This is how the singularity arises at the coinciding point of the Green's function in the renormalization group flows of both U(1) gauge fields and effective vierbein tensor. However, we claim that this UV divergence occurs just in the formal expression because the effective field theory of interacting Dirac fermions has to be defined in a UV cutoff scale $\Lambda$. This UV cutoff scale is proportional to $z_{f} = f d z$, where the renormalization group transformation ends. We recall that $f$ is the number of iteration steps of recursive RG transformations and $d z$ is the renormalization-group transformation scale. $z_{f} = f d z \propto \Lambda$ implies that complete integrations in the frequency-momentum space are performed in the $f-$th iteration step of the renormalization group transformation. In this respect we introduce the UV cutoff proportional to $z_{f}$ directly in the frequency-momentum integral of the Green's function. As a result, such semiclassical solutions of all dynamical fields depend on the UV cutoff in complicated ways, which has to be investigated more seriously.

\subsection{Emergence of an AdS geometry}

Finally, we discuss the emergence of an AdS geometry at the quantum critical point of an interacting Dirac-fermion system, where chiral symmetry breaking is considered. Here, we do not perform our complete analysis for this symmetry breaking transition, which means that (1) more complete mean-field analysis away from the quantum critical point and (2) fluctuation analysis for effective hydrodynamics beyond the mean-field description near the quantum critical point will not be taken into account. However, this simple demonstration shows how the present theoretical framework goes on concretely. Recently, we have shown that the AdS geometry appears at the quantum critical point in the system of interacting scalar fields \cite{Kadanoff_GR_Holography_Kim}. We repeat essentially the same mean-field analysis in the present dual holographic description and confirm that the effective geometry at the quantum critical point of an interacting Dirac-fermion system is given by the AdS metric, indeed.

Turning off current-current and energy-momentum tensor-tensor effective interactions in the present Dirac-fermion system, i.e., $\lambda_{j} = 0$ and $\lambda_{t} = 0$ without additional source fields in Eq. (\ref{Interaction_Dirac_Fermion_System}), we start from the following effective field theory
\bqa && Z = \int D \psi_{\alpha}(x) \exp\Big[ - \int d^{D} x ~ \sqrt{g_{B}(x)} ~ \Big\{ \bar{\psi}_{\alpha}(x) \gamma^{a} e_{B a}^{\mu}(x) \Big( \partial_{\mu} - \frac{i}{4} \omega_{B \mu}^{a'b'}(x) \sigma_{a'b'} \Big) \psi_{\alpha}(x) \nn && + m \bar{\psi}_{\alpha}(x) \psi_{\alpha}(x) + \frac{\lambda_{\chi}}{2 N} \bar{\psi}_{\alpha}(x) \psi_{\alpha}(x) \bar{\psi}_{\beta}(x) \psi_{\beta}(x) \Big\} \Big] , \nonumber \eqa
where spontaneous chiral symmetry breaking is considered instead of current-current effective interactions for simplicity. Here, $e_{B a}^{\mu}(x)$ is a background vielbein tensor introduced just formally. $\alpha$ represents the flavor degeneracy from $1$ to $N$.

Performing the recursive renormalization group transformation discussed in this study, we obtain an effective dual holographic field theory as follows
\bqa && Z = \int D \psi_{\alpha}(x) D \varphi(x,z) D e_{a}^{\mu}(x,z) \delta\Big( e_{a}^{\mu}(x,0) - e_{B a}^{\mu}(x) \Big) \nn && \delta\Big[ \partial_{z} e_{a}^{\mu}(x,z) - \gamma_{a} \Big\{ \gamma^{c} e_{c}^{\mu'}(x,z) \Big( \partial_{\mu'} - \frac{i}{4} \omega_{\mu'}^{c'd'}(x,z) \sigma_{c'd'} \Big) G_{xx'}[\varphi(x,z),e_{a}^{\mu}(x,z)] \Big\}_{x' \rightarrow x} \gamma^{a'} e_{a'}^{\mu}(x,z) \Big] \nn && \exp\Big[ - \int d^{D} x ~ \sqrt{g(x,z_{f})} ~ \Big\{ \bar{\psi}_{\alpha}(x) \gamma^{a} e_{a}^{\mu}(x,z_{f}) \Big( \partial_{\mu} - \frac{i}{4} \omega_{\mu}^{a'b'}(x,z_{f}) \sigma_{a'b'} \Big) \psi_{\alpha}(x) + [m - i \varphi(x,z_{f})] \bar{\psi}_{\alpha}(x) \psi_{\alpha}(x) \Big\} \nn && - N \int d^{D} x ~ \sqrt{g(x,0)} ~ \Big\{ \frac{1}{2 \lambda_{\chi}} [\varphi(x,0)]^{2} \Big\} \nn && - N \int_{0}^{z_{f}} d z ~ \int d^{D} x ~ \sqrt{g(x,z)} ~ \Big\{ \frac{1}{2 \lambda_{\chi}} [\partial_{z} \varphi(x,z)]^{2} + \frac{\mathcal{C}_{\varphi}}{2} g^{\mu\nu}(x,z) [\partial_{\mu} \varphi(x,z)] [\partial_{\nu} \varphi(x,z)] + \mathcal{C}_{\xi} R(x,z) [\varphi(x,z)]^{2} \nn && + \frac{1}{2 \kappa} \Big( R(x,z) - 2 \Lambda \Big) \Big\} \Big] . \nonumber \eqa
Here, $\varphi(x,z)$ is an order parameter field dual to $\bar{\psi}_{\alpha}(x) \psi_{\alpha}(x)$ at $z = z_{f}$ as shown in the IR boundary effective action. The heavy Dirac-fermion propagator is given by \bqa && \Big\{ \gamma^{a} e_{a}^{\mu}(x,z) \partial_{\mu} + \frac{m - i \varphi(x,z)}{\epsilon} \Big\} G_{xx'}[\varphi(x,z),e_{a}^{\mu}(x,z)] = \frac{1}{\sqrt{g(x,z)}} \delta^{(D)}(x-x') , \nonumber \eqa
where $\epsilon$ is a renormalization-group transformation energy scale.

First, we focus on a vacuum solution at zero temperature, where translational invariance is assumed as follows
\bqa && Z = \int D \psi_{\alpha}(x) D \varphi(z) D e_{a}^{\mu}(z) \delta\Big( e_{a}^{\mu}(0) - e_{B a}^{\mu} \Big) \delta\Big[ \partial_{z} e_{a}^{\mu}(z) - \Big\{ \gamma_{a} \Big( \gamma^{c} e_{c}^{\mu'}(z) \partial_{\mu'} G[x-x';\varphi(z),e_{a}^{\mu}(z)] \Big)_{x' \rightarrow x} \gamma^{a'} \Big\} e_{a'}^{\mu}(z) \Big] \nn && \exp\Big[ - \int d^{D} x ~ \sqrt{g(z_{f})} ~ \Big\{ \bar{\psi}_{\alpha}(x) \gamma^{a} e_{a}^{\mu}(z_{f}) \partial_{\mu} \psi_{\alpha}(x) + [m - i \varphi(z_{f})] \bar{\psi}_{\alpha}(x) \psi_{\alpha}(x) \Big\} \nn && - N \int d^{D} x ~ \sqrt{g(0)} ~ \Big\{ \frac{1}{2 \lambda_{\chi}} [\varphi(0)]^{2} \Big\} - N \int_{0}^{z_{f}} d z ~ \int d^{D} x ~ \sqrt{g(z)} ~ \Big\{ \frac{1}{2 \lambda_{\chi}} [\partial_{z} \varphi(z)]^{2} + \mathcal{C}_{\xi} R(z) [\varphi(z)]^{2} + \frac{1}{2 \kappa} \Big( R(z) - 2 \Lambda \Big) \Big\} \Big] . \nonumber \eqa
In other words, the spacetime dependence in $\varphi(x,z)$ and $e_{a}^{\mu}(x,z)$ disappear. Accordingly, the Green's function reads
\bqa && \Big\{ \gamma^{a} e_{a}^{\mu}(z) \partial_{\mu} + \frac{m - i \varphi(z)}{\epsilon} \Big\} G[x-x';\varphi(z),e_{a}^{\mu}(z)] = \frac{1}{\sqrt{g(z)}} \delta^{(D)}(x-x') . \nonumber \eqa
Since the Green's function enjoys the translational symmetry, it can be represented in the energy-momentum space as follows
\bqa && G[k;\varphi(z),e_{a}^{\mu}(z)] = \frac{1}{\sqrt{g(z)}} \frac{- \gamma^{a} e_{a}^{\mu}(z) i k_{\mu} + \frac{m - i \varphi(z)}{\epsilon}}{g^{\mu\nu}(z) k_{\mu} k_{\nu} + \Big(\frac{m - i \varphi(z)}{\epsilon}\Big)^{2}} . \eqa

Introducing this Green's function into the above effective field theory, we obtain
\bqa && \partial_{z} \ln g^{\mu\nu}(z) = 8 \int \frac{d^{D} k}{(2\pi)^{D}} \frac{ g^{\alpha'\beta'}(z) k_{\alpha'} k_{\beta'} }{g^{\alpha\beta}(z) k_{\alpha} k_{\beta} + \Big(\frac{m - i \varphi(z)}{\epsilon}\Big)^{2}} \label{RG_FLOW_METRIC} \eqa
for the emergent metric tensor given by
\bqa && g^{\mu\nu}(z) = e_{a}^{\mu}(z) e_{b}^{\nu}(z) \delta^{ab} \eqa
and
\bqa && \varphi(z) = \frac{\varphi(z_{f})}{\int_{0}^{z_{f}} d y \frac{1}{\sqrt{g(y)}}} \int_{0}^{z} d y \frac{1}{\sqrt{g(y)}} \eqa
for the solution of the chiral-symmetry breaking order parameter, where the UV boundary condition $\varphi(0) = 0$ without an external source field has been utilized. The IR boundary condition is given by
\bqa && \frac{D}{2\lambda_{\chi}} \Bigg( \int \frac{d^{D} k}{(2\pi)^{D}} \frac{g^{\alpha'\beta'}(z_{f}) k_{\alpha'} k_{\beta'}}{g^{\alpha\beta}(z_{f}) k_{\alpha} k_{\beta} + \Big(\frac{m - i \varphi(z_{f})}{\epsilon}\Big)^{2}} \Bigg) i \varphi(z_{f}) = \int \frac{d^{D} k}{(2\pi)^{D}} \frac{ m - i \varphi(z_{f})}{g^{\mu\nu}(z_{f}) k_{\mu} k_{\nu} + \Big(m - i \varphi(z_{f})\Big)^{2}} , \eqa
which results from the Euler-Lagrange equation of the IR boundary effective action $S_{IR}[\varphi(z_{f}),[\partial_{z} \varphi(z)]_{z = z_{f}}]$.

It turns out that the metric tensor gives rise to the AdS geometry when the effective mass of Dirac fermions vanishes, which is essentially the same as that of an interacting scalar field theory \cite{Kadanoff_GR_Holography_Kim}. More precisely, taking $[m - i \varphi(z)]_{z = z_{f}} \longrightarrow 0$ in the renormalization group flow of the metric tensor Eq. (\ref{RG_FLOW_METRIC}), we obtain
\bqa && \partial_{z_{f}} \ln g^{\mu\nu}(z_{f}) = \mathcal{C} , \label{RG_FLOW_METRIC_FP} \eqa
where \bqa && \mathcal{C} = 8 \int \frac{d^{D} k}{(2\pi)^{D}} 1 . \eqa Here, the momentum integral has to be regularized by introducing a cutoff, which is proportional to $z_{f}$ as discussed in the previous subsection. We recall that $z_{f} = f d z$ with the iteration number $f$ of recursive renormalization-group transformations and the renormalization-group energy scale $d z$ defines the integration region of quantum fields (largest momentum where the dynamics of quantum fields are allowed), thus corresponding to the UV cutoff, although it looks like the IR cutoff in the bulk effective action. The fixed-point solution of Eq. (\ref{RG_FLOW_METRIC_FP}) results in the AdS geometry at the quantum critical point. We point out that the full geometry is not the AdS type during the renormalization group flow.

More interestingly, one may extend this analysis to the regime of finite temperatures. The resulting effective field theory is given by
\bqa && Z \approx \int D \psi_{\alpha}(\tau,x) D \varphi(z) D e_{a}^{\tau}(z) D e_{a}^{i}(z) \delta \Big( e_{a}^{\tau}(0) - e_{B a}^{\tau} \Big) \Big( e_{a}^{i}(0) - e_{B a}^{i} \Big) \nn && \delta \Bigg( \partial_{z} \ln e_{a}^{\tau}(z) - \frac{4}{\beta} \sum_{i\omega} \int \frac{d^{d} k}{(2\pi)^{d}} \frac{ e_{b''}^{\tau''}(z) e_{b''}^{\tau''}(z) \omega^{2} }{ e_{b'}^{\tau'}(z) e_{b'}^{\tau'}(z) \omega^{2} + e_{b'}^{i'}(z) e_{b'}^{j'}(z) k_{i'} k_{j'} + \Big(\frac{m - i \varphi(z)}{\epsilon}\Big)^{2}} \Bigg) \nn && \delta \Bigg( \partial_{z} \ln e_{a}^{i}(z) - \frac{4}{\beta} \sum_{i\omega} \int \frac{d^{d} k}{(2\pi)^{d}} \frac{ e_{b''}^{i''}(z) e_{b''}^{j''}(z) k_{i''} k_{j''} }{ e_{b'}^{\tau'}(z) e_{b'}^{\tau'}(z) \omega^{2} + e_{b'}^{i'}(z) e_{b'}^{j'}(z) k_{i'} k_{j'} + \Big(\frac{m - i \varphi(z)}{\epsilon}\Big)^{2}} \Bigg) \nn && \exp\Big[ - \int_{0}^{\beta} d \tau \int d^{d} x ~ \sqrt{g(z_{f})} ~ \Big\{ \bar{\psi}_{\alpha}(\tau,x) \Big( \gamma^{a} e_{a}^{\tau}(z_{f}) \partial_{\tau} + \gamma^{a} e_{a}^{i}(z_{f}) \partial_{i} \Big) \psi_{\alpha}(\tau,x) + [m - i \varphi(z_{f})] \bar{\psi}_{\alpha}(\tau,x) \psi_{\alpha}(\tau,x) \Big\} \nn && - N \int_{0}^{\beta} d \tau \int d^{d} x ~ \frac{\sqrt{g(0)}}{2 \lambda_{\chi}} [\varphi(0)]^{2} \nn && - N \int_{0}^{z_{f}} d z ~ \int_{0}^{\beta} d \tau \int d^{d} x ~ \sqrt{g(z)} ~ \Big\{ \frac{1}{2 \lambda_{\chi}} [\partial_{z} \varphi(z)]^{2} + \mathcal{C}_{\xi} R(z) [\varphi(z)]^{2} + \frac{1}{2 \kappa} \Big( R(z) - 2 \Lambda \Big) \Big\} \Big] . \nonumber \eqa
Here, we separate the vielbein tensor into its time and space component. In addition, all the frequency integrals are replaced with the Matzubara frequency summation, where the time space is $S^{1}$. Then, the question is whether we can have a black hole solution or not. If so, the equation of motion for fluctuations near this finite-temperature mean-field solution may give effective hydrodynamics, as discussed in the previous subsection.

\section{Conclusion}

The holographic duality conjecture describes the holographic liquid in terms of both Maxwell and Einstein coupled equations with an extra dimension. These coupled equations in the extra dimension may be regarded to be a dual description for conserved U(1) currents and energy-momentum tensors in effective hydrodynamics, where the disappearance of quasiparticles would be realized by strong renormalization effects along the extra dimension. In this study, we tried to construct a microscopic foundation for this holographic-liquid phenomenology.

We applied renormalization group transformations a la Polchinski in real space and in a recursive way a la Sung-Sik Lee to interacting Dirac fermions in the presence of both background metric or vierbein and external electromagnetic U(1) gauge fields. Besides detailed mathematical derivations, it turns out that both external source fields minimally coupled to conserved currents are uplifted to be dynamical and are forced to renormalization-group flow in the emergent extra-dimensional space. In other words, global symmetries of quantum field theories turn into gauge symmetries through the renormalization group flows of the boundary source fields in the extra dimensional space. As a result, we obtain an effective Einstein-Maxwell theory in $(D+1)$ spacetime dimensions from the quantum field theory of interacting Dirac fermions in $D$ spacetime dimensions. IR boundary conditions self-consistently determined within this effective field theory serve as renormalization group $\beta-$functions for all types of interaction vertices including field renormalization.

It is natural to expect that our microscopically derived Einstein-Maxwell theory would show the holographic-liquid phenomenology. However, we point out that the renormalization group flows of all dynamical fields are nonlinearly intertwined beyond the conventional holographic setup. In this respect it is necessary to look at possible solutions more carefully and compare these results with recent experiments near future. In addition, we emphasize that our derivation has not been performed in a covariant way. It is also important to develop a covariant formulation in our opinion.

\section*{Acknowledgement}

K.-S. Kim was supported by the Ministry of Education, Science, and Technology (No. 2011-0030046) of the National Research Foundation of Korea (NRF) and by TJ Park Science Fellowship of the POSCO TJ Park Foundation. K.-S. Kim appreciates fruitful discussions with Shinsei Ryu and his hospitality during the sabbatical leave. K.-S. Kim thanks Sung-Sik Lee for his comments on the manuscript.

\end{document}